\documentclass[aps,preprint,nofootinbib]{revtex4-1}
\usepackage{graphicx}
\usepackage[latin1]{inputenc}
\usepackage{epsfig}
\usepackage{amsmath}
\usepackage{amsfonts}
\usepackage{amssymb}
\usepackage{perpage} 
\usepackage{slashed}
\usepackage[final]{feynmp}
\usepackage{hyperref}

\makeatletter
\def\endfmffile{%
	\fmfcmd{\p@rcent\space the end.^^J%
			end.^^J%
			endinput;}%
	\if@fmfio
		\immediate\closeout\@outfmf
	\fi
	\ifnum\pdfshellescape=\@ne
		\immediate\write18{mpost \thefmffile}%
	\fi}
\makeatother
\MakePerPage{footnote}
\def\lsim{\:\raisebox{-0.5ex}{$\stackrel{\textstyle<}{\sim}$}\:}
\def\gsim{\:\raisebox{-0.5ex}{$\stackrel{\textstyle>}{\sim}$}\:}
\newcommand{\Lagr}{\mathcal{L}}
\newcommand{\wt}[1]{\widetilde{#1}}
\begin{document}
\begin{titlepage}
\begin{flushright}
\vspace*{-0.5cm}
{\small
TUM-HEP-1032/15
}
\end{flushright}
\vspace*{0.5cm}
\begin{center}
  \begin{LARGE}
    \begin{bf}
The Flavor of the Composite Twin Higgs
   \end{bf}
  \end{LARGE}
\end{center}
\vspace{0.1cm}
\begin{center}
\begin{large}
{\bf Csaba Cs\'aki$^a$, Michael Geller$^b$, Ofri Telem$^b$, and Andreas Weiler$^c$}
\end{large}
  \vspace{0.5cm}
  \begin{it}

\begin{small}
$^{(a)}$ Department of Physics, LEPP, Cornell University, Ithaca, NY 14853, USA
 \vspace{0.2cm}\\
$^{(b)}$Physics Department, Technion - Israel Institute of Technology, Haifa 32000, Israel
\vspace{0.2cm}\\
$^{(c)}$Physik Department T75, James-Franck-Stra{\ss}e 1, Technische Universit\"at M\"unchen, 85748 Garching, Germany
\vspace{0.2cm}\\
 \end{small}
\end{it}

{\tt csaki@cornell.edu, mic.geller@gmail.com, t10ofrit@gmail.com, andreas.weiler@tum.de}

\end{center}


\begin{center}
{\large Abstract}
\end{center}

\noindent The assumption of anarchic quark flavor puts serious stress on composite Higgs models: flavor bounds imply a tuning of a few per-mille (at best) in the Higgs potential. Composite twin Higgs (CTH) models significantly reduce this tension by opening up a new region of parameter space, obtained by raising the coupling among the composites close to the strong coupling limit $g_* \sim  4\pi$, thereby raising the scale of composites to around 10 TeV. This does not lead to large tuning in the Higgs potential since the leading quantum corrections are canceled by the twin partners (rather than the composites). We survey
the leading flavor bounds on the CTH, which correspond to tree-level $\Delta F=2$ four-Fermi operators from Kaluza-Klein (KK) Z exchange in the kaon system and 1-loop corrections from KK fermions to the electric dipole moment of the neutron. We provide a parametric estimate for these bounds and also perform a numeric scan of the parameter space using the complete calculation for both quantities. The results confirm our expectation that CTH models accommodate anarchic flavor significantly better than regular composite Higgs (CH) models. Our conclusions apply both to the identical and fraternal twin cases.

\bigskip

\end{titlepage}

\section{Introduction}

The search for traditional top partners responsible for canceling the leading contribution to the Higgs boson mass has so far come up empty handed, leading to bounds of order 700 GeV for stops and vectorlike top partners \cite{Aad:2015kqa,CMS:2013tda}. These bounds are starting to put significant pressure on supersymmetric and composite Higgs models. An alternative approach to solving the hierarchy problem is to consider models with top partners that are not charged under ordinary QCD, but rather under a mirror QCD group, related by a discrete symmetry to the Standard Model (SM) \cite{Chacko:2005pe,Chacko:2005un,Curtin:2015fna,Csaki:2015fba,Curtin:2015bka,Craig:2013fga,Craig:2015pha,Craig:2014aea,Craig:2014roa,GT,Low:2015nqa,Barbieri:2015lqa}. In this case the direct bounds on the top partners disappear, and natural models with small fine tuning are still viable. The best known example of such models is the Twin Higgs (TH)\cite{Chacko:2005pe}, in which the entire SM gauge group is doubled, and the twin sector is related by a softly broken $Z_2$ symmetry to the visible SM. In these models the one-loop quadratic divergences are automatically canceled, and the hierarchy problem is postponed until the cutoff scale of the theory of order 5-10 TeV, where a UV completion of the model becomes necessary. In particular, theories of flavor, whose scale is generally well above the multi-TeV, can only be incorporated into the TH once its UV completion is specified.

The simplest UV completion of the TH is by extending it to a composite Higgs (CH) model \cite{GT,Low:2015nqa,Barbieri:2015lqa}, with the scale of composite resonances as high as ${\cal O}(10)$ TeV. This could either be a warped extra dimensional (ED) model \cite{GT} or a 4D composite Higgs model with partial compositeness for the fermions \cite{Low:2015nqa,Barbieri:2015lqa}. In this paper we will be using the warped extra dimensional language, though every result can be restated in terms of the  corresponding 4D CH model.

The main feature of this UV completion is the appearance of additional gauge and fermion partners (the KK modes in the extra dimensional language). However, unlike traditional CH models \cite{GeorgiKaplan,ACG,LH,Agashe:2004rs,Agashe:2006at,Contino:2006qr,Contino:2006nn,SILH,Panico:2015jxa,Panico:2012uw,CHreviews}, these KK gauge and fermion partners are not the states responsible for the cancellation of the 1-loop divergences of the Higgs mass - that role is played by the twin partners: the twin top, twin W and Z, etc. The KK modes are simply there to UV complete the theory. Within this framework it is now possible to examine the question of the flavor hierarchy and flavor constraints on TH models. Of particular interest is the fact that warped ED models actually provide a natural framework for explaining the origin of the observed flavor hierarchies \cite{HuberShafi,Agashe:2004cp,CFW,rattazzi}. The appearance of small Yukawa couplings in this scenario is due to exponentially small overlaps of extra dimensional wave functions. This RS-flavor mechanism also incorporates a natural suppression of flavor changing neutral currents (FCNC) called the RS-GIM mechanism: the same wave function suppressions appearing in the Yukawa couplings will also suppress the flavor changing operators. While RS-flavor is a very intriguing possibility, a detailed examination of the flavor constraints shows that there is a significant tension left between the KK scale needed for natural electroweak symmetry breaking (EWSB) and that needed to evade flavor constraints. See refs.~\cite{Agashe:2004cp,CFW,rattazzi,Agashe:2008uz,Casagrande:2008hr,Albrecht:2009xr,Blanke:2008zb,Blanke:2008yr} for the discussion of $\Delta F=2$ bounds and refs.~\cite{ABP,GIP,CGTT,Beneke1,rattazzi,Konig:2014iqa,MR,Beneke2,btosgamma} for the leading dipole bounds in warped models. Generically in CH models, $M_{KK} <1$ TeV is required for a fully natural Higgs mass, while the flavor bounds require $M_{KK} > 10-20$ TeV. To this end, flavor symmetries have been introduced to reconcile flavor bounds with naturalness \cite{Cacciapaglia:2007fw,Redi:2011zi,CFW2,Csaki:2009wc,Fitzpatrick:2007sa,Barbieri:2012uh,Redi:2013eaa}.

The purpose of this paper is to reexamine these flavor bounds in the context of CTH models. While the form of the expressions for the flavor bounds in CTH models are practically identical to those in warped ED, the KK modes no longer play the role of the top partners, allowing us to explore a different region of parameter space. In particular, in CH models the tuning of the Higgs mass grows with $(g_* f)^2$, where $g_*$ is the interaction strength of the KK modes, and $f$ is the global symmetry breaking scale. Thus, one cannot raise $g_*$ without increasing the fine-tuning in the Higgs sector. In contrast, the fine tuning in the CTH model grows as $f^2$ and is insensitive to $g_*$ (since the cancellation is achieved by the twin partners).
As we will show in this paper, the tension in the CTH model between flavor constraints and Higgs sector tuning is significantly reduced for larger values of $g_*$.
When $g_* \sim 2\pi~(4\pi)$, we obtain a scenario with $f\sim 3~(2)$ TeV, where all flavor constraints are obeyed and the tuning is at a percent level. One interesting consequence of raising $g_*$ is that the leading contribution to $\Delta F=2$ FCNC is no longer from KK gluon exchange, but rather from KK Z exchange. In this work we are agnostic to the description of the light mirror quarks and our analysis is relevant whether the low energy theory is fraternal~\cite{Craig:2015pha} or identical~\cite{Chacko:2005pe} twin Higgs. We focus on flavor constraints from the quark sector, and leave the lepton sector for a future study.

The paper is organized as follows: in Sec.~\ref{sec:estimates} we provide a brief overview of our main results; in Sec.~\ref{sec:CTH} we define the CTH model and calculate the Higgs potential and the tuning. In Sec.~\ref{sec:flavor} we go over the calculation of the flavor bounds and in Sec.~\ref{sec:scan} we give the details of the numerical scan. Our conclusions are contained in Sec.~\ref{sec:Conclusions}. A series of appendices detail the construction of fermions in the CTH model (App.~\ref{Appendix:reps}), the fraternal CTH model (App.~\ref{Appendix:FBC}), the evaluation of the Higgs potential (App.~\ref{Appendix:Higgs}), the $Z_2$ breaking via hypercharge (App.~\ref{Appendix:Z2b}), the tuning in CH models (App.~\ref{Appendix:TuneCH}), and the loop functions relevant for the dipole calculations (App.~\ref{Appendix:dipole}).

\section{Overview of flavor bounds\label{sec:estimates}}

Before presenting the detailed calculations of the flavor bounds in the CTH model, we present a summary of the expected results based on simple estimates. We then emphasize the improvement of the fine tuning in the Higgs sector in the CTH model vs. standard composite Higgs models, in the presence of flavor bounds. The details of the CTH model will be reviewed in Sec.~\ref{sec:CTH}, and will be used in the full evaluation of the flavor bounds in Secs.~\ref{sec:flavor}-\ref{sec:scan}.

The key parameters of the model are the global symmetry breaking scale $f$, and the (dimensionless) interaction strength $g_*$ of the composite states (KK modes). The KK mass is related to these parameters via
\begin{equation}
M_{KK}= g_* f \ .
\end{equation}
There is an additional coupling $g_{s*}$ which measures the interaction strength of the KK gluon. Since the models under consideration are 5D warped theories with the Higgs arising as a scalar component of the 5D gauge field, there is no actual Yukawa coupling parameter $Y$ in these models. Instead the Yukawa couplings arise from the 5D gauge interactions, and will also be proportional to some dimensionless boundary localized mixing parameters denoted by $\tilde{m}_{u,d}$. The full meaning of these parameters can be obtained by the expression of the standard model masses
\begin{eqnarray}
m_u &\sim& \frac{g_* v}{2\sqrt{2}} f_Q \tilde{m}_u f_{-u}\nonumber \\
m_d &\sim& \frac{g_* v}{2\sqrt{2}} f_Q \tilde{m}_d f_{-d}
\end{eqnarray}
where the functions $f_{Q,-u,-d}$ are the standard RS zero mode wave functions evaluated at the IR brane. The usual 5D RS Yukawa couplings can thus be identified as $Y_{u,d} = g_{*}\frac{\tilde{m}_{u,d}}{2}$. Note that there is also kinetic mixing in these 5D models among the matter fields, which gives additional contributions to the expressions above. However it does not play a role in the simple estimate and we ignore it for now. An estimate for the $C^4_K$ and $C^5_K$ coefficients of the $\Delta F=2$ color-singlet L-R 4-Fermi operator relevant for the kaon sector are given by~\cite{CFW}:
\begin{equation}
\begin{array}{c}
C^4_K \sim \frac{1}{M_{KK}^2} \frac{g_{s*}^2}{g_{*}^2} \frac{8 m_d m_s}{v^2} \frac{1+\tilde{m}_d^2}{\tilde{m}_d^2}\\
C^5_K \sim \frac{1}{3} \left( 4 \frac{g_{*}^2}{g_{s*}^2} - 1\right) C^4_K,
\label{DelF2}\end{array}
\end{equation}
with $C^4_K$ mediated by the KK gluon and $C^5_K$ by both the KK gluon and KK Z. The KK Z contribution to $C^5_K$ dominates at higher values of $g_*$. Note that the basic structure of this expression does not depend on whether one considers a CTH model or a standard CH model in warped space. Clearly there are various ${\cal O}(1)$ factors showing up due to the enlarged group structure. These will be included in the full scan of Sec.~\ref{sec:scan}, and we ignore them for now. Assuming $g_{s*} =6$ (i.e., no boundary kinetic terms for the $SU(3)$) , $\tilde{m}_d \ll 1$ and substituting $m_d \sim 3$ MeV, ${m}_s = 47$ MeV for the quark masses at 3 TeV, we have
\begin{equation}
\begin{array}{c}
C^4_{K,estimate} = \frac{1}{\left(1.6\times 10 ^5 { \ \rm TeV}\right)^2}\left(\frac{106~{\rm TeV}}{g_*^2 f \tilde{m}_d}\right)^2 \\
C^5_{K,estimate} = \frac{1}{\left(1.4\times 10 ^5 { \ \rm TeV}\right)^2}\left(\frac{106~{\rm TeV}}{g_*^2 f \tilde{m}_d}\right)^2
\frac{1}{4}\left[ {\left( \frac{g_{*}}{3} \right)}^2 - 1\right]
\end{array}
\end{equation}
From the bound on the imaginary part of these operators we obtain the $\Delta F=2$ bound on the parameters of the model
\begin{equation}
g_*  f \tilde{m}_d > 106~{\rm TeV},
\label{KKbarlimit}
\end{equation}
for $g_* < 6.7$ where KK gluon exchange and the resulting bound from $C^4_K$ dominates, and
\begin{equation}
g_* f \tilde{m}_d >  17.7~{\rm TeV},
\label{KKbarlimitC5}
\end{equation}
for $g_* > 6.7$ where KK Z exchange and $C^5_K$ dominates.
The most stringent dipole bound on these models arises from contributions to the neutron EDM:
\begin{equation}
\frac{c}{8\pi^2f^2} m_d \overline{d}_L \sigma^{\mu \nu} e F_{\mu\nu}  d_R +\frac{\tilde{c}}{8\pi^2f^2} m_d \overline{d}_L \sigma^{\mu \nu} g_s G_{\mu\nu}  d_R.
\end{equation}
A simple estimate for the one-loop contribution of the KK fermions is given by
\begin{equation}
c\sim \tilde{c}\sim \frac{1}{g^2_* m_d} \frac{v}{\sqrt{2}} f_Q Y_dY_d^\dagger Y_d f_{-d} \sim \frac{1}{g^2_*} (Y^2) = \frac{\tilde{m}_d^2 }{4}
\end{equation}
up to ${\cal O}(1)$ factors. These factors are calculated in Sec.~\ref{sec:flavor}, and the corresponding limit is
\begin{equation}
\frac{f}{\tilde{m}_d}> 2.85 \ {\rm TeV}\ .
\label{EDMlimit}
\end{equation}
The two limits (\ref{KKbarlimit})-(\ref{EDMlimit}) together imply the bound
 \begin{equation}
  g_*f>\rm max \left(1,\sqrt{\frac{g_*}{6.7}}\right)17.3\  {\rm TeV}\ .
\label{eq:mainbound}
\end{equation}
When this bound is saturated we have
\begin{equation}
\tilde{m}_d= 0.48\frac{4\pi}{g_{*}} \rm max \left(1,\sqrt{\frac{g_*}{6.7}}\right).
\end{equation}
\\
In ordinary composite Higgs models the tuning is given by
\begin{equation}\label{eq:Tuning_CH}
\Delta_{CH} \sim a {\left(\frac{g_* f}{v}\right)}^2,
\end{equation}
where $a$ is an $\mathcal{O}$(1) constant that depends on the particular CH model (for example, $a=0.35$ in the model of \cite{CFW} and \cite{Carena:2007ua}, see  Appendix~\ref{Appendix:TuneCH}).
In other words, the flavor bounds constrain the tuning directly, to the sub per-mille level. This estimate might be somewhat pessimistic for the composite Higgs models, since it assumes a universal scale for both the composite vector bosons and the composite fermions. Relaxing this condition, the estimates of Eq.~(\ref{DelF2}) become
\begin{equation}
\begin{array}{c}
C^4_{K,estimate} = \frac{1}{{\left(g_*^V f\right)}^2} \left(\frac{g_{s*}}{g^F_{*}}\right)^2 \frac{8 m_d m_s}{v^2} \frac{1+\tilde{m}_d^2}{\tilde{m}_d^2}\\
C^5_{K,estimate} = \frac{1}{3} \left( 4 \left(\frac{g^V_{*}}{g_{s*}}\right)^2 - 1\right) C^4_K,
\end{array}
\end{equation}
where $g_*^V f $ and $g_*^F f$ are the scales of the vector and the fermion excitations. Of these two scales, it is only the scale of the fermion excitation that dominates in the tuning. The limit is then
 \begin{equation}
 \sqrt{ g^F_*}f>\rm max \left(\sqrt{\frac 1{g^V_*}},\sqrt{\frac 1{6.7}}\right)17.3\  {\rm TeV}\,
\end{equation}
implying in this case a tuning of a few per-mille at best. In comparison, the tuning in CTH models is only linked to $f$ and not $g_*$:
\begin{equation}
\Delta_{CTH} \sim 1 \times \frac{f^2}{v^2}.
\end{equation}
For larger values of $g_*$, Eq.~(\ref{eq:mainbound}) is satisfied with a smaller $f$, and consequently, smaller tuning. In this regime the dominant $\Delta F=2$ bound comes from the KK Z mediated $C^5_K$. Of course in this extreme case of $g_* \sim 4\pi$ the 5D description of the CTH model itself becomes strongly coupled. For a more realistic value of $g_* \sim 2 \pi$, $f \sim 3~{\rm TeV}$ and the tuning is at the percent level. To this extent, the idea of anarchic flavor can be revived in the framework of composite twin Higgs. These considerations apply both to identical and fraternal~\cite{Craig:2015pha} versions of the CTH.

\section{The Composite Twin Higgs\label{sec:CTH}}

In this section we define our benchmark composite twin Higgs model, whose flavor structure will be studied in detail in the upcoming sections of this paper. This model is realized in 5D anti-de Sitter (AdS) space with gauge-Higgs unification (GHU). We parametrize the 5D AdS metric in the standard form:
\begin{equation}
ds^2=\left(\frac{R}{z}\right)^2\left(dx_\mu dx_\nu \eta^{\mu\nu}-dz^2\right),
\end{equation}
where $R$ is the AdS curvature and $R \leq z \leq R'$ is the coordinate of the extra dimension.  The UV brane is located at $z=R$ and the IR brane at $z=R^\prime$, with the hierarchy $R^\prime/R\sim 10^{16}$ between the weak and Planck scales.

The gauge symmetry in the bulk is
\begin{equation}
SO(8) \times \left(SU(3)_c \times U(1)_X\right)^{SM} \times \left(SU(3)_c \times U(1)_X\right)^{m} \times Z_2^{SM \leftrightarrow m},
\end{equation}
where the SO(8) contains the SM electroweak SU(2)$_L$ as well as the SU(2)$_R$ needed for custodial symmetry, and their twin partners
\begin{equation}
\left(SU(2)_L \times SU(2)_R\right)^{SM} \times \left(SU(2)_L \times SU(2)_R\right)^{m}\subseteq SO(8).
\end{equation}
The bulk symmetry is broken on the UV and IR branes to
\begin{eqnarray}
IR&:&~SO(7) \times \left(SU(3)_c \times U(1)_X\right)^{SM} \times \left(SU(3)_c \times U(1)_X\right)^{m} \times Z_2^{SM \leftrightarrow m} \\
UV&:&~\left(SU(3)_c\times SU(2)_L \times U(1)_Y\right)^{SM}\times \left(SU(3)_c\times SU(2)_L \times U(1)_Y\right)^{m} \times Z_2^{SM \leftrightarrow m},
\end{eqnarray}
by imposing Dirichlet (-) boundary conditions for the gauge bosons corresponding to broken generators.
The UV brane boundary conditions ensure that the surviving low-energy gauge symmetry is SM$\times $SM$_{twin}$, while the IR breaking SO(8)$\to$SO(7) ensures the emergence of the pseudo Nambu-Goldstone boson (pNGB) Higgs necessary for the twin Higgs mechanism.
 The hypercharge is as usual $Y=X+T^3_R$, and the mirror hypercharge is defined analogously as  $Y^m=X^m+T^{3m}_R$. The seven broken generators in the coset $SO(8)/SO(7)$ are denoted $T^{i8}$, $i=1,\ldots ,7$. The $A^5_i$ components of the gauge bosons corresponding to the these generators get IR Neumann (+) boundary conditions. Of these only  $A^5_{1,\ldots ,4}$ have also UV Neumann boundary conditions - the four zero modes corresponding to the pNGB Higgs doublet,  in the ${\bf 4}$ of ${\left(SU(2)_L \times SU(2)_R\right)}^{SM}$. A Coleman-Weinberg potential for these zero modes arises through loops of SM and mirror gauge bosons and fermions, resulting in the SM electroweak symmetry breaking.

The $A^5_{1,\ldots ,4}$ enter the equations of motion (EOM) of the gauge and fermion fields through the Wilson line between the two branes:
\begin{equation}
\Omega(R,R^\prime)= e^{ig_5 T^{i8} \int{ dz A^5_i}} = e^{\frac{i\sqrt{2} \ T^{i8}h_i}{f}},
\end{equation}
where $h_i (x) \equiv A^5_i (x)$, while the scale $f$ and coupling $g_*$ (introduced in Sec.~\ref{sec:estimates}) are formally defined by
\begin{equation}
f \equiv \frac{2}{g_*R^\prime}~,~ g_* \equiv \frac{g_5}{ \sqrt{R}}.
\end{equation}
$f$ can be thought of as the vacuum expectation value (VEV) corresponding to the $SO(8)/SO(7)$ breaking, and $g_*$ is the dimensionless bulk gauge coupling of $SO(8)$ characterizing the interaction strength of the KK modes. The KK scale is defined as:
\begin{equation}
M_{KK}\equiv \frac{2}{R^\prime} = g_* f
\end{equation}
the latter relation was already used in Sec.~\ref{sec:estimates}.
Using gauge transformations  we can always bring $\Omega$ into the form:
\begin{equation}
\Omega = \left(
\begin{array}{cccccccc}
1  & 0 & 0 & 0 & 0 & 0 & 0 & 0\\
0  & 1 & 0 & 0 & 0 & 0 & 0 & 0\\
0  & 0 & 1 & 0 & 0 & 0 & 0 & 0\\
0  & 0 & 0 & \cos(\frac{v}{f}) & 0 & 0 & 0 & \sin(\frac{v}{f})\\
0  & 0 & 0 & 0 & 1 & 0 & 0 & 0\\
0  & 0 & 0 & 0 & 0 & 1 & 0 & 0\\
0  & 0 & 0 & 0 & 0 & 0 & 1 & 0\\
0  & 0 & 0 & -\sin(\frac{v}{f}) & 0 & 0 & 0 & \cos(\frac{v}{f})
\end{array}
\right)\label{Omega}
\end{equation}
\subsection{The Quark Sector}
The SM and mirror quarks in our model are embedded in bulk multiplets. Specifically, $q_L$ is in the $\bf 8$ of $SO(8)$, $d_R$ is in the anti-symmetric representation $\bf 28$, while $u_R$ is an SO(8) singlet:
\begin{eqnarray}
\Psi_{\mathbf{8}}&=&\left( \begin{array}{cl} q_L(+,+)& T^3_R=-1/2\\ ...(-,+) \end{array} \right)_{\left({\bf 3},\frac{2}{3},{\bf 1},0\right)}
 ~  \Psi_{\mathbf{1}}=\left( \begin{array}{c} u_R(+,+) \end{array} \right)_{\left({\bf 3},\frac{2}{3},{\bf 1},0\right)}\\
  \Psi_{\mathbf{28}}&=&\left( \begin{array}{c}
d_R(+,+)~T^3_R=-1\\ ...(-,+)\end{array}  \right)_{\left({\bf 3},\frac{2}{3},{\bf 1},0\right)}
\end{eqnarray}
Here  $(+/-)$ denote the UV/IR boundary conditions, and the subscripts are the representations under $\left(SU(3)_c \times U(1)_X\right)^{SM} \times \left(SU(3)_c \times U(1)_X\right)^m$. Note that the above boundary conditions are for the left component of $\Psi_{\mathbf{8}}$ and the right component of $\Psi_{\mathbf{1}},\Psi_{\mathbf{28}}$. It is evident from the boundary conditions that $q_L$ and $t_R$ are the only zero modes in the SM sector. All other components, denoted here by $\ldots$ , do not have zero modes. These components are explicitly listed  in Appendix~\ref{Appendix:reps}.
The twin partners are embedded in the $Z^{SM\leftrightarrow m}_2$ duals of the above multiplets, in the following way:
\begin{eqnarray}
\Psi^m_{\mathbf{8}}&=&\left( \begin{array}{cl} q^m_L(+,+)& T^{3m}_R=-1/2\\ ...(-,+) \end{array} \right)_{\left({\bf 1},0,{\bf 3},\frac{2}{3}\right)}
 ~  \Psi^m_{\mathbf{1}}=\left( \begin{array}{c} u^m_R(+,+) \end{array} \right)_{\left({\bf 1},0,{\bf 3},\frac{2}{3}\right)}\\
  \Psi^m_{\mathbf{28}}&=&\left( \begin{array}{c}
d^m_R(+,+)~T^{3m}_R=-1\\ ...(-,+)\end{array}  \right)_{\left({\bf 1},0,{\bf 3},\frac{2}{3}\right)}
\end{eqnarray}
The alternative fraternal CTH realization of the fermion sector (where only the third generation fermions have zero modes in the mirror sector) is detailed in App.~\ref{Appendix:FBC}. Note that the leading flavor constraints will be identical in both cases.

Under the IR $SO(7)$ symmetry, the bulk multiplets decompose as
\begin{eqnarray}
\Psi_{\mathbf{8}} &\to&  \Psi^{\mathbf{1}}_{\mathbf{8}L}, \Psi^\mathbf{7}_{\mathbf{8}L} \nonumber\\
\Psi_{\mathbf{28}} &\to&  \Psi^{\mathbf{7}}_{\mathbf{28}R}, \Psi^{\mathbf{21}}_{\mathbf{28}R}\nonumber\\
\Psi_{\mathbf{1}} &\to&  t_R,
\end{eqnarray}
For each component we only show the chirality that has (+) boundary conditions (b.c.) on the IR brane. In the mirror sector, due to the $Z_2$ on the IR brane we have:
\begin{eqnarray}
\Psi^m_{\mathbf{8}} &\to&  \Psi^{\mathbf{1}m}_{\mathbf{8}L}, \Psi^{\mathbf{7}m}_{\mathbf{8}L}\nonumber\\
\Psi^m_{\mathbf{28}} &\to&  \Psi^{\mathbf{7}m}_{\mathbf{28}R}, \Psi^{\mathbf{21}m}_{\mathbf{28}R}\nonumber\\
\Psi^m_{\mathbf{1}} &\to&  t^m_R,
\end{eqnarray}
 The IR symmetry allows for the following mass terms on the IR brane:
\begin{equation}
\Lagr_{IR}=-\left(\frac{R}{R'}\right)^4 \left[ \wt{m}_u \left( \Psi^\mathbf{1}_{\mathbf{8}L} t_R+\Psi^{\mathbf{1}m}_{\mathbf{8}L} t^m_R \right)+\wt{m}_d \left( \Psi^\mathbf{7}_{\mathbf{8}L} \Psi^\mathbf{7}_{\mathbf{28}R}+\Psi^{\mathbf{7}m}_{\mathbf{8}L}\Psi^{\mathbf{7}m}_{\mathbf{28}R}  \right) \right].
\end{equation}
The $\wt{m}_{u,d}$ are dimensionless IR mass parameter matrices. Note, that the IR Lagrangian has to be invariant under the $Z_2$ on the IR brane and so the SM and mirror multiplets share the same IR masses. These masses are generically $3\times3$ anarchic matrices, unless a flavor symmetry is postulated on th IR brane. With this choice of representations and boundary conditions, the lightest states in $q_L,u_R,d_R$ are zero modes prior to EWSB. The effect of the IR masses is to rotate the zero modes among the bulk multiplets, and as a result some zero modes will appear in more than one bulk multiplet. Specifically the left handed $q_L$ lives in $\Psi_8$ and in $\Psi_{28}$, the right-handed $u_R$ lives in $\Psi_{1}$ and in $\Psi_8$ and the right-handed $d_R$ lives only in $\Psi_{28}$.
This results in kinetic mixing for each zero mode in the "bulk basis" - the basis in which the bulk masses are diagonal. This kinetic mixing can be parameterized by three Hermitian matrices $K_q,K_u,K_d$ so that the kinetic term is $\bar{\Psi}K\slashed{D}\Psi$~\cite{CFW}:
\begin{eqnarray}
K_q&=&1+f_q\wt{m}_d f_d^{-2}\wt{m}^\dagger_df_q \nonumber\\
K_u&=&1+f_{-u}\wt{m}_u f_{-q}^{-2}\wt{m}^\dagger_uf_{-u} \nonumber\\
K_d&=&1 \label{kineticmix}
\end{eqnarray}
where $f_c$ is the standard RS flavor function
\begin{equation}
f_c=\sqrt{\frac{1-2c}{1-\left(\frac{R'}{R}\right)^{2c-1}}} \label{ffunction}
\end{equation}

These fermions get mass due to the VEV of $A^5$. This VEV enters the bulk EOM through the covariant derivative of each multiplet in the bulk Lagrangian. To find the masses in the fermion KK tower in the presence of the $A^5$ VEV, we use an "auxiliary" fermion field for each multiplet:

\begin{equation}
\Psi^\prime(h)=\Omega \Psi.
\end{equation}
This auxiliary field has the Wilson lines rotated away from its EOM and thus satisfies the same bulk EOM as $\Psi$ in the $A^5 \to 0$ case, and $\Omega$ defined in Eq.~(\ref{Omega}), however the b.c. of the auxiliary fields will now contain $\Omega$. This rotation now mixes the different zero modes appearing in the same multiplet (due to the IR localized masses) generating mass terms for the fermions after EWSB. Using the same basis as in Eq.~(\ref{kineticmix}), the mass terms are:
\begin{eqnarray}
 m_u &=& \frac{g_* v}{2\sqrt{2}} f_Q \tilde{m}_u f_{-u}\nonumber \\
m_d &=& \frac{g_* v}{2\sqrt{2}} f_Q \tilde{m}_d f_{-d} \label{quark_masses}
\end{eqnarray}

\subsection{The top sector and the Higgs potential}

To find the Higgs potential, we find the profile of the top and mirror top multiplets by solving the bulk EOM and imposing the b.c.  The solution is expressed in the form of the spectral function $\rho(p^2)=\det (-1+\frac{m^2}{p^2})$, whose zeroes in $p$ are exactly the (VEV dependent) masses forming the full the KK tower:
\begin{eqnarray}
\rho_t(p^2) &=& 1 + f_t(p^2) \ \sin^2\left(\frac{h}{f}\right) \\
\rho_{tm}(p^2) &=& 1 + f_t(p^2) \ \cos^2\left(\frac{h}{f}\right)
\end{eqnarray}
where
\begin{equation}
f_t=-\frac{\frac{1}{2} C_{-1} \left(\frac{R'}{R}\right)^{2c_u-2c_q}\wt{m}^2_u}{\left(C_{-8}S_{1} + C_{-1}S_{8}\left(\frac{R'}{R}\right)^{2c_u-2c_q} \wt{m}^2_u\right) S_{-8}}
\end{equation}
and $C_{\pm i} \equiv C_{\pm c_i}(R',p)$, $S_{\pm i} \equiv S_{\pm c_i}(R',p)$ are two linear combinations of Bessel functions that solve the bulk equation and are most convenient to use with $+/-$ boundary conditions (for the definition see App.~\ref{Appendix:Higgs}).

The contribution of the top/mirror top sector to the Coleman-Weinberg potential of the Higgs is then given by
\begin{equation}
V_{eff}(h)=\frac{-4N_c}{(4\pi)^2}\int_0^\infty dp\,  p^3 \log(\rho_t[-p^2]\rho_{tm}[-p^2]).
\end{equation}

As explained in~\cite{CFW} (and also in App.~\ref{Appendix:Higgs}), we can approximate the contribution $V_{eff}$ by:
\begin{equation}
V_{eff}(h)=-\alpha_2 \sin^2\frac{h}{f} - \frac{\alpha}{2} \sin^4 \frac{h}{f} - \alpha_2 \cos^2\frac{h}{f} - \frac{\alpha}{2} \cos^4 \frac{h}{f}
\label{CTHpotentialFull}
\end{equation}
For all our results we use the full calculation for these terms that appears in App.~\ref{Appendix:Higgs}. Here we give the NDA estimates:
\begin{eqnarray}
\alpha&\sim& -\frac{3}{64\pi^2}y_t^4f^4 \left(1+2\log\frac{2g_*^2}{y_t^2}\right)\\
\alpha_2&\sim& \frac{3}{16 \pi^2}y_t^2 g_*^2 f^4
\end{eqnarray}

These are the standard loop generated terms in the composite Higgs potential, where $\alpha$ is the logarithmically divergent term and $\alpha_2$ is the quadratically divergent term.

In the Higgs potential of Eq.~(\ref{CTHpotentialFull}), the $\sin \frac{h}{f}$ terms are generated by the top and the $\cos \frac{h}{f}$ are generated by the mirror top. The quadratically divergent contribution sums up to a constant piece in the potential independent of the Higgs field. The only remaining piece in the potential is:
\begin{equation}
V_{eff}(h)=-\alpha \sin^2 \frac{h}{f}\cos^2 \frac{h}{f}
\label{CTHpotential}
\end{equation}

Due to the $Z_2$ invariance in the top sector, the Higgs potential in Eq.~(\ref{CTHpotential}) has a minimum for $v=\frac{f}{\sqrt{2}}$. To obtain a realistic $v\ll f$, an additional (at this point unspecified) $Z_2$ breaking contribution must be present which we parametrize as
\begin{equation}
V(h)=-\alpha \sin^2\frac{h}{f}\cos^2\frac{h}{f}+ \beta \sin^2\frac{h}{f}
\end{equation}
 For any value of $f$ we can now calculate $\alpha$ and $\beta$ that produce the right Higgs mass and VEV:
\begin{eqnarray}
\alpha &=& \alpha_0 \frac{1}{1- \frac{v^2}{f^2}}\\[1em]
\beta &=& \alpha_0 \frac{1}{1- \frac{v^2}{f^2}} \left(1- 2\frac{v^2}{f^2}\right)
\label{alphabetaf}\end{eqnarray}
where for convenience we define $\alpha_0$ as the value of $\alpha$ and $\beta$ for $f \to \infty$
\begin{equation}
\alpha_0 =   \frac{f^4 m^2_h}{8v^2}
\end{equation}

\begin{figure}[h]
\includegraphics[scale=0.4]{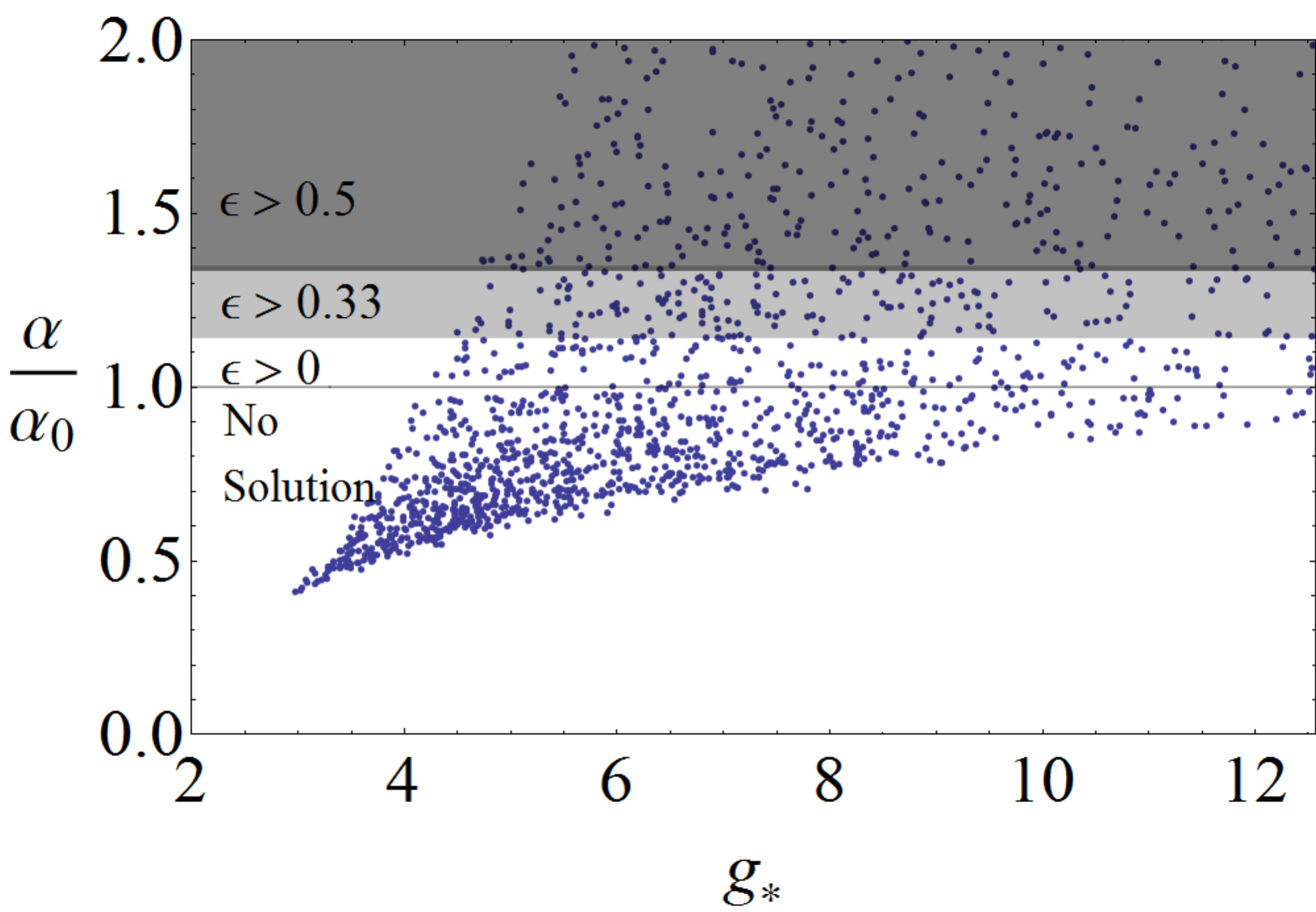}
\caption{A scatter plot of $\frac{\alpha}{\alpha_0}$ as a function of $g_*$ chosen to get the correct top mass. The chosen parameters are $0<c_q<0.35$, $-0.5<c_u<-0.3$, $0.3<\tilde{m}_u<1$ and $R'/R=10^{16}$. $\alpha>\alpha_0$ is required to get the correct Higgs mass, and for $f\to \infty$ the required $\alpha$ approaches $\alpha_0$. We mark on the plot the regions of $\alpha$ that are required for $\epsilon\equiv \frac{v}{f}>\frac{1}{3}$ and for $\epsilon>\frac{1}{2}$.
\label{Potential}}
\end{figure}

In Fig.~\ref{Potential} we plot $\alpha/\alpha_0$ as a function of $g_*$ for a variety of input parameters in the 5D model. We find that the points in the 5D model naturally populate the various values of $\frac{\alpha}{\alpha_0}$ for large enough values of $g_*$. For $\alpha < \alpha_0$ there is no solution that gives the right EWSB, and similarly for $\beta > \alpha_0$.

If the $Z_2$ is exact in the gauge sector, then the contribution of gauge boson loops to the Higgs potential is negligible for our purposes:
\begin{equation}
\alpha^{gauge} \sim \frac{9}{256\pi^2}g^4f^4 \lsim \frac{\alpha }{20}.
\end{equation}

\begin{figure}[h]
\includegraphics[scale=0.4]{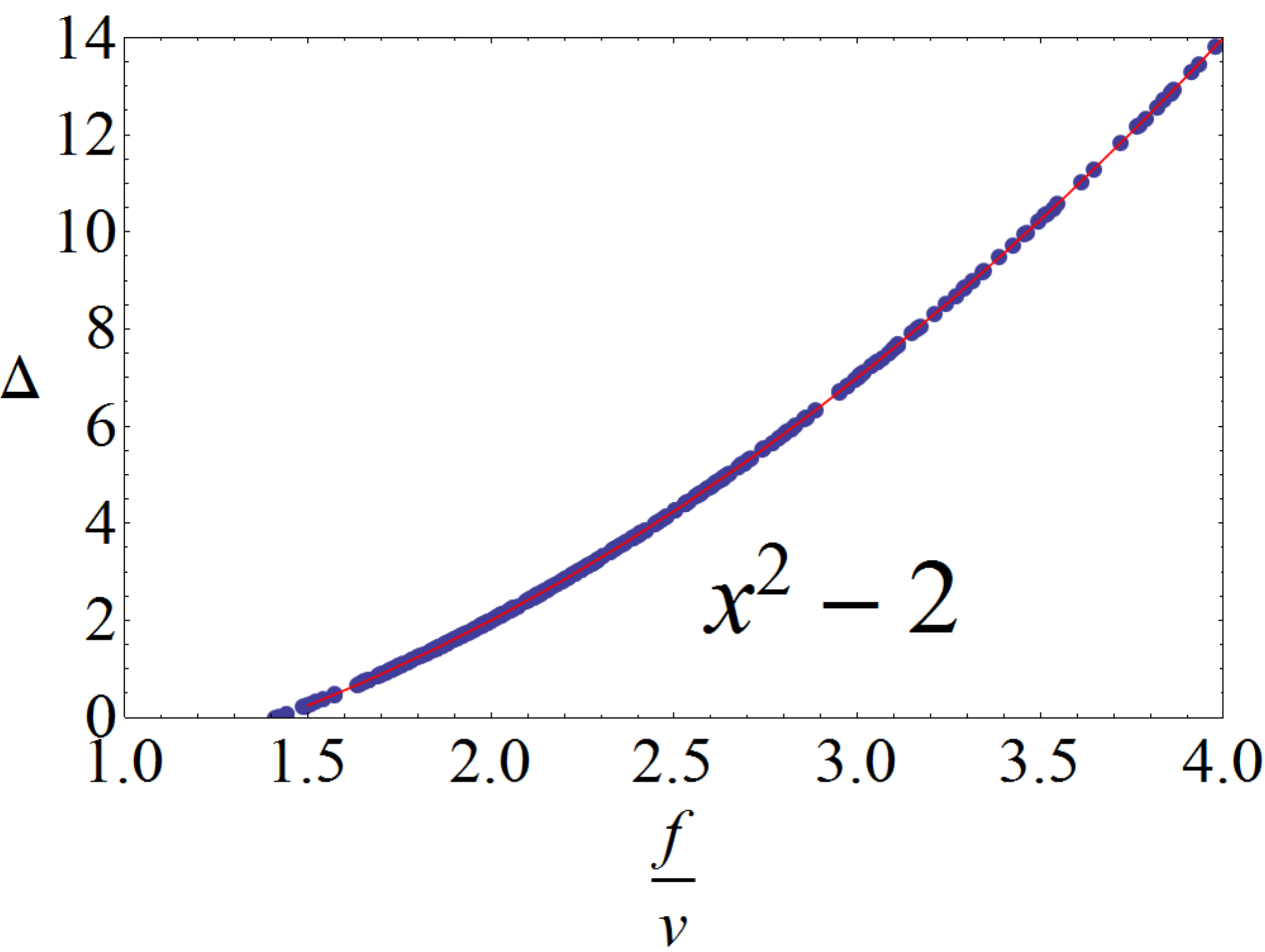}
\caption{The overall tuning in the CTH as function of $\frac{f}{v}$. The red line is a quadratic fit to the tuning: $\Delta = \frac{f^2}{v^2}-2$, which for large f can be approximated as $\frac{f^2}{v^2}$.
\label{PlotTuning}}
\end{figure}

In App.~\ref{Appendix:Z2b} we give an example of a $Z_2$ breaking contribution to the Higgs potential that comes from a difference in the bulk gauge couplings of $U(1)_X$ and $U(1)_X^m$ breaking the $Z_2$, such that for every value of $f$ there is a $g_X^m<g_X$ that gives the right $\beta$. The precise origin of the $Z_2$ breaking has no bearing on the flavor observables therefore we remain agnostic to it.
The tuning is calculated using the Barbieri-Giudice measure:
\begin{equation}
\Delta = \max \left|\frac{d\log v}{d\log p_i}\right|= \max\left(\Delta_{\alpha},\Delta_{\beta}\right) = \frac{1}{2} \left(\frac{f^2}{2v^2}-1\right)\max \left|\frac{d\log \alpha, \beta}{d\log p_i}\right|
\label{tuning}\end{equation}
where $p_i$ are all the parameters of the 5D theory. The tuning for different points in the parameter space of the model with the $Z_2$ contribution of App.~\ref{Appendix:Z2b} is plotted in Fig.~\ref{PlotTuning}. The red line is a quadratic fit to the tuning given by:
\begin{equation}
\Delta_{CTH}= \frac{f^2}{v^2}-2.
\end{equation}

We note that the tuning only depends on $f$ and not on $g_*$ which is essentially a free parameter. This is the major difference compared to the CH, where the tuning $\Delta_{CH}\sim g_*^2$, and thus $g_*$ can be raised only at the expense of increased tuning.

\section{Anarchic Flavor Bounds \label{sec:flavor}}

The main purpose of this paper is to establish the basic flavor bounds on the CTH with anarchic flavor. To this end we focus on the flavor physics in the quark sector, while the lepton sector is left for a future study.

The flavor constraints on CH have been studied extensively in \cite{Agashe:2004cp,CFW,rattazzi,Agashe:2008uz,Azatov:2014lha,Casagrande:2008hr,Fitzpatrick:2007sa,Cacciapaglia:2007fw,Blanke:2008zb,Blanke:2008yr,Albrecht:2009xr,Redi:2011zi,Csaki:2009wc,Barbieri:2012uh,Azatov:2014lha}, providing a limit on $M_{KK}$ for anarchic flavor. In CH models, the tuning in the CH is proportional to $M^2_{KK}$, and so any bound on $M_{KK}$ is also a lower bound on the tuning.\footnote{The tuning in CH depends on the representations in the top sector \cite{Panico:2012uw}. In App.~\ref{Appendix:TuneCH} we calculate the tuning in the model with two adjoints and a fundamental of SO$(5)$, which has "double tuning" according to the definitions of~\cite{Panico:2012uw}. }  The strongest constraint in the quark sector  $M_{KK}\gsim 20$ TeV arises from $\epsilon_K$, implying a per-mille level tuning at the least. We note that this bound may be lower in the case when the masses of the vector excitations are parametrically lower than the mass of the KK tops that enters the Higgs potential.

In the CTH, the flavor bounds are parametrically the same as in the CH. The Higgs potential, however, is only logarithmically dependent on $M_{KK}$, and so $g_*$ is essentially a free parameter. As we will see, the lower bound on the tuning in the CTH implied by the flavor bounds will get weaker for larger values of $g_*$ as it is raised toward the strong coupling limit $g_*\lsim4\pi$.

We start with the flavor structure of the CTH, working in the "bulk basis" - the basis in which the bulk masses are diagonal.
The anarchic mass matrices are given by:
\begin{eqnarray}
m^{ij}_u &=&\left( \frac{g_* v}{2\sqrt{2}} F_Q \tilde{M}_u F_{-u}\nonumber\right)^{ij} \\
m^{ij}_d &=& \left(\frac{g_* v}{2\sqrt{2}} F_Q \tilde{M}_d f_{-d}\right)^{ij}
\end{eqnarray}
where we have simply extended Eq.~(\ref{quark_masses}) to all three generations. The kinetic terms are:
\begin{eqnarray}
K^{ij}_q&=&\delta^{ij}+\left(F_q\tilde{M}_d F_d^{-2}\tilde{M}^\dagger_d F_q \right)^{ij} \nonumber\\
K^{ij}_u&=&\delta^{ij}+\left(F_{-u}\tilde{M}_u F_{-q}^{-2}\tilde{M}^\dagger_u F_{-u} \right)^{ij} \nonumber\\
K^{ij}_d&=&\delta^{ij}
\end{eqnarray} Here $\tilde{M}^{ij}_{u/d}$ are the anarchic $3\times 3$ IR mass matrices, and $F_c$ is a diagonal matrix whose diagonal elements are the RS flavor functions $f_c$ (see Eq.~(\ref{ffunction})):
\begin{equation}
F_u = \text{Diag}\left(f_u, f_c ,f_t\right)~,~F_d = \text{Diag}\left(f_d , f_s , f_b\right)~,~F_q = \text{Diag}\left(f_{q_1},f_{q_2},f_{q_3}\right)
\end{equation}
 To find the physical quark masses we must first diagonalize the kinetic terms by rotating the quarks with the Hermitian matrices $H_q$ and $H_u$ ($H_d$ is already the identity matrix due the structure of the b.c. ):
\begin{equation}
q_L\to H_q q_L~,~u_R \to H_u u_R
\end{equation}
The rotation to the mass basis is then via the usual unitary matrices $U_L,U_R,D_L,D_R$. The resulting diagonal mass matrices are:
\begin{eqnarray}
M_u&=&\frac{g_* v}{2\sqrt{2}} U^\dagger_L H_q F_Q \tilde{M}_u F_{-u}\nonumber H_u U_R \nonumber \\
M_d&=&\frac{g_* v}{2\sqrt{2}} D^\dagger_L H_q F_Q \tilde{M}_d F_{-d} D_R
\end{eqnarray}
As a result, all of the quark couplings are rotated as well, yielding the source of flavor violations.

We now present the relevant constraints for the $\Delta F =2$ and the dipole operators.

\subsection{$\Delta F =2 $}

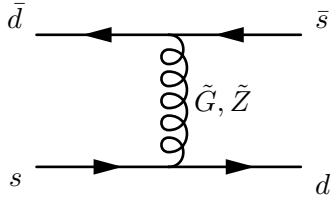
\begin{figure}[h]
\vspace*{1cm}
\begin{center}
	    \begin{fmffile}{kaon}
	        \begin{fmfgraph*}(100,50)
                 \fmfstraight
	        \fmfleft{i1,i2}
            \fmflabel{q}{i1}
	        \fmfright{o1,o2}
                 \fmflabel{$s$}{i1}
                 \fmflabel{$\bar{d}$}{i2}
                 \fmflabel{$d$}{o1}
                 \fmflabel{$\bar{s}$}{o2}
                 \fmf{fermion,tension=1}{i1,v1,o1}
                 \fmf{fermion,tension=1}{o2,v2,i2}
                 \fmf{gluon,tension=0,label=$\ \tilde{G},,\tilde{Z}$}{v1,v2}
	        \end{fmfgraph*}
	    \end{fmffile}
\end{center}
\caption{Tree level $\Delta F =2$ flavor violation in the kaon system, mediated by KK gluon/KK Z.  \label{delFeq2}}
\end{figure}

$\Delta F =2 $ processes are mediated mainly by KK gluons and KK Z bosons (see Fig.~\ref{delFeq2}). The KK gluon couplings in the mass basis are:
\begin{eqnarray}
g^d_L &=& D_L^\dagger H_q \left(g^{dL}_8(G)+F_q \tilde{M_d}F_d^{-2}g^{dL}_{28}(G) \tilde{M_d}^\dagger F_q\right) H_q D_L  \\
g^u_L &=& U_L^\dagger H_q \left(g^{uL}_8(G)+F_q \tilde{M_d}F_d^{-2}g^{uL}_{28}(G) \tilde{M_d}^\dagger F_q\right) H_q U_L  \\
g^d_{R} &=& D_R^\dagger g^{dR}_{28}(G) D_R  \\
g^u_{R} &=& U_R^\dagger H_u \left(g^{uR}_{1}(G) +F_{-d} \tilde{M_u}F_{-q}^{-2}g^{uR}_8(G) \tilde{M_u}^\dagger F_{-d}\right) H_u U_R
\end{eqnarray}
where $g^x_\Psi(V)$ is a diagonal matrix that gives the wavefunction overlap of the vector boson V with the $x$ component of the $\Psi$ multiplet:
\begin{equation}
g^x_\Psi(G) = g_{s*}F^{+/-}(c_\Psi)~,~\Psi = \Psi_8,\Psi_{28},\Psi_1 ~,~x=u_L,u_R,d_L,d_R,
\end{equation}
where $F^{+/-}(c_\Psi)$ is the function that gives the overlap of the gauge KK state with (+/-,+) b.c. and a fermion with bulk mass $c_\Psi$.
For the KKZ, there are two distinct KK states: $Z_{KK}$ -- the KK mode of the SM $Z$ boson, with $(+,+)$ b.c. and $B'$  -- the gauge boson corresponding to the neutral broken generator in $SU(2)_R \times U(1)_x/U(1)_Y$, with $(-,+)$ b.c. These states mix after EWSB~\cite{Agashe:2004cp,Albrecht:2009xr} with mixing angle $\xi$, and the mass eigenstates are denoted as $Z_H$ and $Z'$. The wavefunction overlaps for these states are:
\begin{eqnarray}
g^x_\Psi(Z_H) &=& \cos \xi \frac{g_*}{\cos \theta_W}F^+(c_\Psi) a^x_\Psi(Z) +  \sin \xi \frac{g_*}{\cos \phi}F^-(c_\Psi) a^x_\Psi(B')\\
g^x_\Psi(Z') &=& -\sin \xi \frac{g_*}{\cos \theta_W}F^+(c_\Psi) a^x_\Psi(Z) +  \cos \xi \frac{g_*}{\cos \phi}F^-(c_\Psi) a^x_\Psi(B'),
\end{eqnarray}
where $a^x_\Psi(Z)$ and $a^x_\Psi(B')$ are functions of the quantum numbers of the $x$ component in $\Psi$ given by:
\begin{eqnarray}
a^x_\Psi(Z) \Psi_x &=&(T^3_L-Q \ \mbox{sin}^2 \theta_W) \Psi_x \\
a^x_\Psi(B') \Psi_x &=& (T^3_R-(T^3_R+X) \tan \theta_W) \Psi_x.
\end{eqnarray}

With these couplings we can now calculate the $\Delta F=2$ constraints on $g_*$ and $f$. We focus on the kaon system, specifically on the $\Delta F =2$ operators with the strongest experimental constraints:
\begin{eqnarray}
  \text{Im}(C^4_K) &(&\bar{s}^{\alpha}_L d^{\alpha}_R)(\bar{s}^{\beta}_R d^{\beta}_L)~,~\Lambda_F > 1.6\times 10^5~\text{TeV}\\
  \text{Im}(C^5_K) &(&\bar{s}^{\alpha}_L d^{\beta}_R)(\bar{s}^{\beta}_R d^{\alpha}_L)~,~\Lambda_F > 1.4\times 10^5~\text{TeV}
\end{eqnarray}
The KK gluon contribution to $C^4_K,C^5_K$ can be calculated:
\begin{equation}
\text{Im}(C^4_K)=-\text{Im}(3C^5_k)= \text{Im}(g^{s12}_{L} g^{s21}_{R}) \sim \frac{1}{f^2}\frac{g_{s*}^2}{g_*^4} \frac{1}{\tilde{m}^2} \frac{8m_d m_s}{v^2}\label{KKGkaon}
\end{equation}
While the KKZ contribution is:
\begin{equation}
C^4_K=0~,C^5_k = 2 \, \text{Im}(g^{Z_H12}_{L} g^{Z_H12}_{R}+g^{Z'12}_{L} g^{Z'21}_{R}) \sim \frac{4}{3f^2}\frac{1}{g_*^2} \frac{1}{\tilde{m}^2} \frac{8m_d m_s}{v^2}\label{KKZkaon}
\end{equation}

From Eqs.~(\ref{KKGkaon})-(\ref{KKZkaon}) we learn that both contributions decouple with large $g_*$, but the KKZ contribution dominates at large $g_*$ since it scales with $\frac{1}{g_*^2}$ compared to the $\frac{1}{g_*^4}$ dependence of the KK gluon contribution. In our calculations, we assume no boundary kinetic terms for the $SU(3)$ so that,
\begin{equation}
 g_{s*}=g_s\log\frac{R'}{R}\approx 6
\end{equation}
 In the numerical scan we calculate $C^4_k$ using the full expressions, and the estimate for the bound is given by:
\begin{equation}
\frac{g_*f}{\tilde{m_d}} \gsim 17.7~{\rm TeV}
\end{equation}
for $g_*>6.7$. $\tilde{m}_d$ is an average IR mass.

\subsection{Dipole Operators}

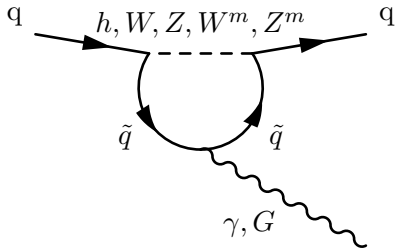
\begin{figure}[h]
\begin{center}
\vspace*{1cm}
	    \begin{fmffile}{dipole}
	        \begin{fmfgraph*}(125,80)
                 \fmfstraight
	        \fmfleft{i2,i1}
            \fmflabel{q}{i1}
            \fmflabel{q}{o1}
	        \fmfright{o2,o1}
                 \fmf{fermion,tension=1}{i1,v1}
                 \fmf{fermion,tension=1}{v2,o1}
                 \fmf{dashes,tension=1,label=$h,,W,,Z,,W^m,,Z^m$}{v2,v1}
                 \fmf{fermion,right=0.6,tension=0.2,label=$\tilde{q}$}{v1,v3,v2}
                \fmf{phantom,right=1,tension=0.2}{i2,v3,o2}
                 \fmf{photon,tension=0,label=$\gamma,,G$}{o2,v3}
	        \end{fmfgraph*}
	    \end{fmffile}
\end{center}
\caption{The main contribution in the CTH model to the dipole operators $\bar{q}q\gamma,\bar{q}qG$, with KK fermions in the loop together with a Higgs, EW or mirror EW boson.\label{dipole}}
\end{figure}

The second type of constraints arise from loop induced dipole operators\footnote{We thank David Straub for discussions and for providing us with an updated version of~\cite{Konig:2014iqa}.}. We follow here the method of~\cite{Konig:2014iqa}. The main contribution to these constraints is the KK fermion loop (see Fig.~\ref{dipole}). For the dipole calculation, we work in the approximation where we keep only one KK level for the fermions, and only the zero modes for the gauge bosons. With this approximation the fermion fields are:
\begin{eqnarray}
\text{Zero mode level}&:& \Psi^{Q8}_{L},\Psi^{u1}_{R},\Psi^{d28}_{R} \nonumber\\
\text{First KK level}&:& \Psi^{KK8}_{L,R},\Psi^{KK1}_{L,R},\Psi^{KK28}_{L,R} \nonumber
\end{eqnarray}
In this notation, the zero mode multiplets include only the components with $(+,+)$ b.c. , and the rest of the components are set to zero; for instance, the only non-zero component in $\Psi^{Q8}_L$ is $Q_L$. The mass terms are:
\begin{eqnarray}
L&=&m_{Q} \bar{\Psi}^{KK8}_L \Psi^{KK8}_R+m_{D} \bar{\Psi}^{KK28}_L \Psi^{KK28}_R+m_{U}\bar{\Psi}^{KK1}_L \Psi^{KK1}_R + \nonumber\\
&+&\frac{\tilde{m}_d}{R'}\left[\left(\bar{\Psi}^{KK8}_L+f_{c_8}\bar{\Psi}^{Q8}_L\right) \left(\Psi^{KK28}_R+f_{c_{28}}\Psi^{d28}_R\right) \Omega+\Omega \bar{\Psi}^{KK28}_L\Psi^{KK8}_R\right]+\nonumber\\
 &+&\frac{\tilde{m}_u}{R'}\left[\left(\bar{\Psi}^{KK8}_L+f_{c_8}\bar{\Psi}^{Q8}_L\right) \Omega \left(\Psi^{KK1}_R +f_{c_1} \Psi^{u1}_{R}\right) \Omega+\bar{\Psi}^{KK1}_L \Omega \Psi^{KK8}_R \right]+{\rm h.c.}\ ,
\end{eqnarray}
where we have assumed that all the components in the first KK-level of each multiplet have a common
mass, denoted by $m_Q,m_D$ and $m_U$ for $\Psi^{KK8}, \Psi^{KK28}$ and $\Psi^{KK1}$, respectively. We note that there is a slight difference in the KK masses of states with $(+,+)$ and $(-,+)$ b.c. In our approximation we neglect this difference, and set $m_Q = m_U = m_D = g_* f$.  Rotating to the mass basis, we find the interactions with the gauge bosons of the form
\begin{equation}
i\bar{q} \left(V^{q\Psi}_{XL}P_L+V^{q\Psi}_{XR}P_R  \right)X \Psi~,~X=h,\slashed W,\slashed Z,\slashed W^m,\slashed Z^m,
\end{equation}
where $q$ is a zero mode quark and $\Psi$ is a first KK mode quark. All the dipole operators can be calculated from the couplings $V^{q\Psi}_X$. The strongest bound is the electric dipole moment of the neutron, where the parton level contributions are:
\begin{equation}
\frac{c}{8\pi^2f^2}m_d \overline{d}_L \sigma^{\mu \nu} e F_{\mu\nu}  d_R+\frac{\tilde{c}}{8\pi^2f^2} m_d \overline{d}_L \sigma^{\mu \nu} g_s G_{\mu\nu}  d_R.
\end{equation}
The coefficients $c,\tilde{c}$ can be calculated \cite{Konig:2014iqa}:
\begin{equation}
c = f^2\sum\limits_{\Psi,X} \frac{m_\Psi}{m_d m_X^2}V^{d\Psi}_{XR}V^{d\Psi*}_{XL}L^\Psi_X~,~\tilde{c} =f^2 \sum\limits_{\Psi,X}\frac{m_\Psi}{m_d m_X^2} V^{d\Psi}_{XR}V^{d\Psi*}_{XL}\tilde{L}^\Psi_X.
\end{equation}
Here, $L_X,\tilde{L}_X$ are the loop functions defined in \cite{Konig:2014iqa}, and are given in Appendix~\ref{Appendix:dipole}. Our calculation gives to the leading order in $f_8,f_{28},f_1$ and $v$:
\begin{eqnarray}
c&=&\frac{1}{4 m_d}\frac{1}{g^2_*} \frac{v}{\sqrt{2}} D_L^\dagger H_d F_Q Y_d Y^\dagger_d  Y_d F_{-d}D_R\\
\tilde{c}&=&\frac{9}{4 m_d}\frac{1}{g^2_*} \frac{v}{\sqrt{2}} D_L^\dagger H_d F_Q Y_d Y_d^\dagger  Y_d F_{-d}D_R,
\end{eqnarray}
where $Y_d = \frac{g_*}{2}\tilde{m_d}$. The experimental bounds are estimated to be~\cite{Konig:2014iqa}
\begin{equation}
\label{dipoleEstimate}
\frac{f}{\sqrt{c}}>3.11~\text{TeV}~,~\frac{f}{\sqrt{\tilde{c}}}>3.79~\text{TeV}.
\end{equation}
We note that this estimation is based on QCD sum rules, evaluating both contributions to the neutron EDM individually, and may underestimate the theoretical uncertainty. We can now establish the previously quoted estimate of the resulting bound:
\begin{equation}
\frac{f}{\tilde{m_d}}>2.85\  {\rm TeV},
\end{equation}
where $\tilde{m}_d$ is an average IR mass.

\subsection{Tree level $\Delta F=1$ constraints}

An additional form of constraints are tree-level $\Delta F=1$ constraints due to the off-diagonal couplings of the $Z$ boson. These constraints are typically satisfied in CH when $f\gsim 1$ TeV, and only get weaker with higher $g_*$. Here we give an example of $t\to cZ$. For high $g_*$, The main contribution is $t_L \to c_L Z$, as the right handed chirality is protected by the custodial symmetry that protects $Z \to b_L b_L$. The $P_{LR}$ symmetry \cite{Agashe:2006at,Contino:2006qr} is satisfied for right handed up quarks $T^3_L = T^3_R =0$, and for left handed down quarks $T^3_L=T^3_R = -\frac{1}{2}$. Thus, all non universal contribution to $Z$ coupling to these states has to come from mixing with the other chirality. The branching ratio for left handed tops can be estimated:
\begin{equation}
Br(t\to c Z) \sim 1 \times 10^{-6} \frac{(1~{\rm TeV})^4}{f^4} \frac{1}{g^2_*}
\end{equation}
For $g_* \to 4\pi$, this branching fraction is far below the LHC reach even for $f$ below 1 TeV. These constraints are subdominant and we do not include them in our numerical scan.
We note here that similar contributions to $Z^m$ mediated FCNC processes are absent due to the exact $P^m_{LR}$ symmetry for all the SM quarks ($T^{3m}_L=T^{3m}_R=0$).

\section{Numerical scan\label{sec:scan}}

In this section we establish the validity of our estimates by performing a numerical scan of the flavor bounds. The full expressions for the magnitudes of the flavor violating processes together with the previously obtained simple estimates are:
\begin{eqnarray}
\text{Im}(C^4_K)&=& \text{Im}(g^{s12}_{L} g^{s21}_{R}) \sim \frac{1}{\left(1.6\times 10 ^5 { \ \rm TeV}\right)^2}\left(\frac{106~{\rm TeV}}{g_*^2 f \tilde{m}_d}\right)^2\nonumber\\
\text{Im}(C^5_K) &=&\text{Im}\left(-\frac{C^4_K}{3}+ 2g^{Z_H12}_{L} g^{Z_H12}_{R}+2g^{Z'12}_{L} g^{Z'21}_{R}\right) \sim \nonumber\\
&\sim&  \frac{1}{\left(1.4\times 10 ^5 { \ \rm TeV}\right)^2}\left(\frac{106~{\rm TeV}}{g_*^2 f \tilde{m}_d}\right)^2
\frac{1}{4}\left[ {\left( \frac{g_{*}}{3} \right)}^2 - 1\right],
\end{eqnarray}
for the $\Delta F=2 $ operators, and
\begin{eqnarray}
c&=&\frac{1}{4 m_d}\frac{1}{g^2_*} \frac{v}{\sqrt{2}} D_L^\dagger H_d F_Q Y_d Y^\dagger_d  Y_d F_{-d}D_R\sim \frac{1}{4}\frac{\tilde{m}^2_d}{4} \\
\tilde{c}&=&\frac{9}{4 m_d}\frac{1}{g^2_*} \frac{v}{\sqrt{2}} D_L^\dagger H_d F_Q Y_d Y_d^\dagger  Y_d F_{-d}D_R\sim \frac{9}{4}\frac{\tilde{m}^2_d}{4},
\end{eqnarray}
for the coefficients of the operators contributing to the neutron EDM.

To test our estimates, we generated sets of bulk and IR masses that reproduce the flavor texture of the SM and give the right EWSB. We then calculated the above expressions with these sets of parameters and compared to the estimates. We assumed that the IR masses form a $3\times 3$ matrix, with eigenvalues randomly distributed and a random unitary rotation from the eigenbasis. The comparison with the estimate is performed by taking the $\tilde{m}_d$ in the estimates as the mean of the probability distribution of the eigenvalues of the $3\times 3$ matrix.

The procedure is then the following. We start by generating 7000 sets of 5D parameters in the top sector that give a realistic Higgs potential, as explained in App.~\ref{Appendix:Higgs}. These parameters are defined by the bulk
masses of the third generation quarks $c_{q_3}, c_{u_3}, c_{d_3}$, the third generation IR boundary mass parameter $\tilde{m}_{u_3}$ and $g_*, R, R',f$. For each set of parameters we then follow the following steps

  $\bullet$  Fix the remaining bulk masses such that the physical quark masses and CKM angles will be naively reproduced.

$\bullet$ Choose random complex boundary mass matrices $\tilde{m}_u, \tilde{m}_d$, with one of the eigenvalues of  $\tilde{m}_u$ being $\tilde{m}_{u_3}$. The rest of the eigenvalues are chosen uniformly between $\left[\frac{1}{3}\tilde{m}_{u_3},\frac{5}{3}\tilde{m}_{u_3}\right]$. These are the anarchic IR mass matrices.

   $ \bullet$ Calculate the full mass matrix and perform a $\chi^2$-fit for the six mass parameters, the three CKM angles, and the Jarlskog invariant. If the fit is reasonable ($\chi^2 < \xi$) keep these parameters, otherwise throw it away.

    $\bullet$ Using this procedure we generate 100 sets of flavor parameters for each 5D parameter point that reproduces correct EWSB. For each set we calculate magnitudes of all the relevant flavor violation processes.

\begin{figure}[h]
\includegraphics[scale=0.4]{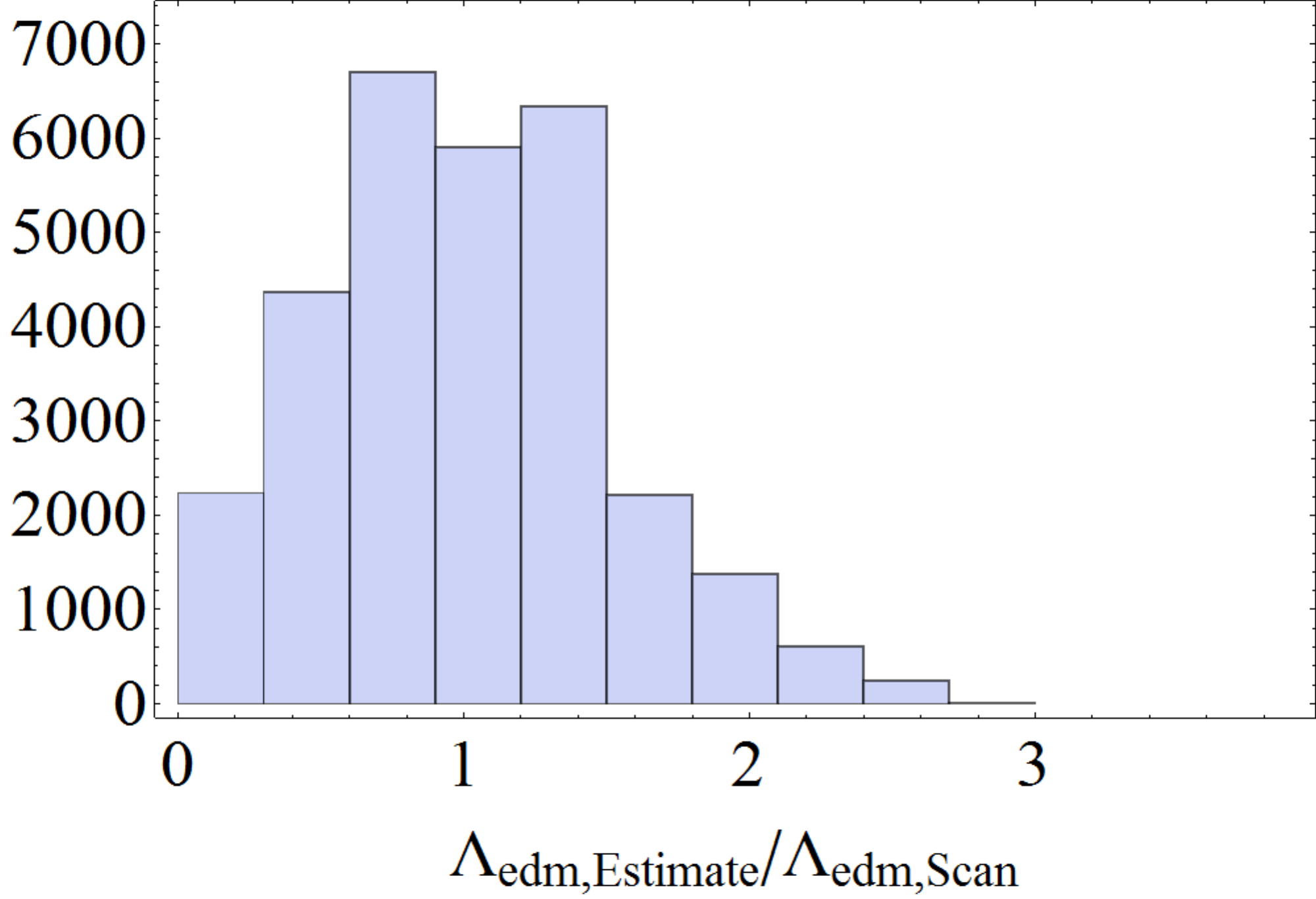}
\includegraphics[scale=0.4]{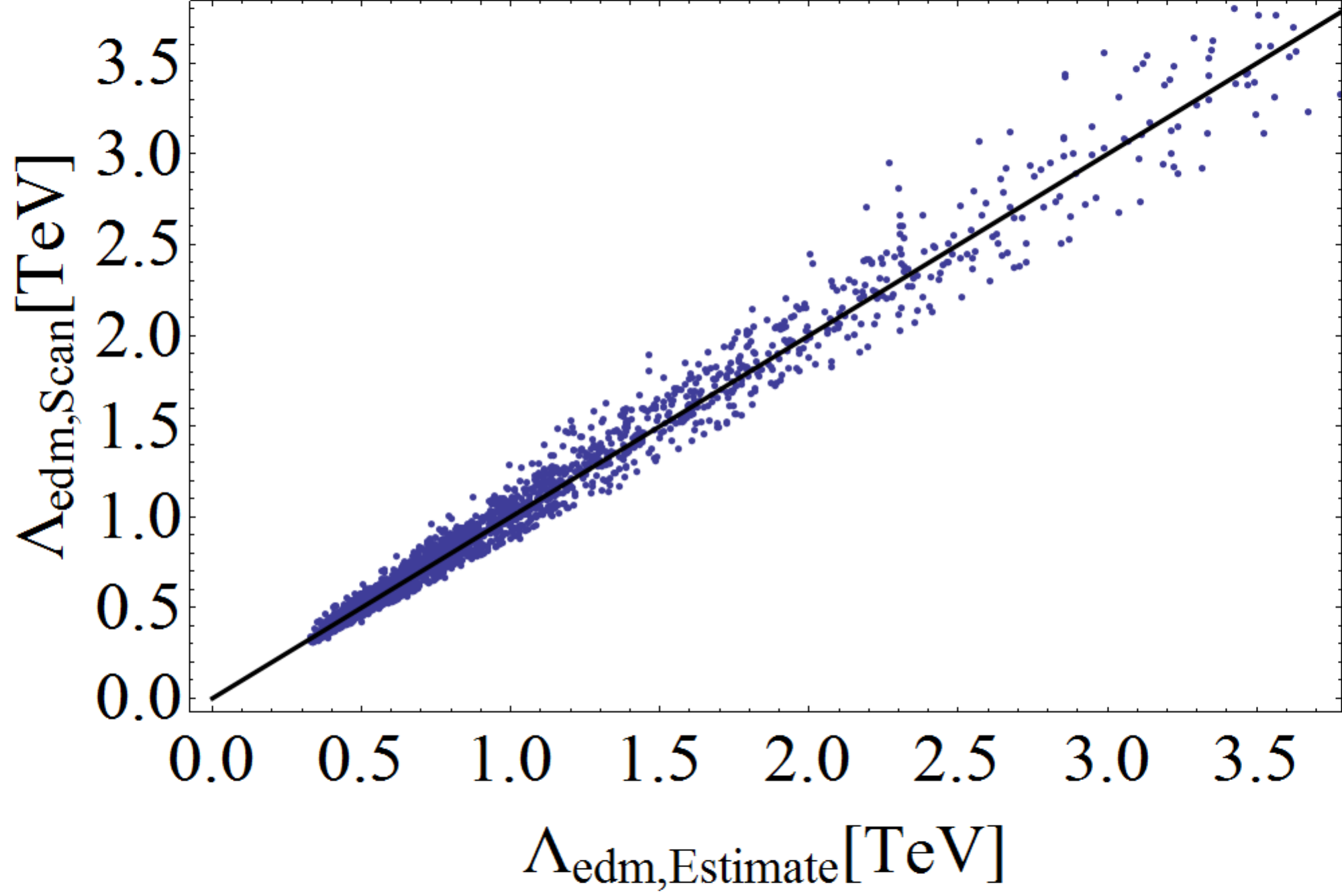}
\caption{A scatter plot of the suppression scale of the gluon contribution to neutron EDM: $\Lambda_{EDM}\equiv \frac{f}{\sqrt{\tilde{c}}}$. The left plot is the histogram of the distribution of $\frac{\Lambda_{edm,\text{estimate}}}{\Lambda_{edm,\text{scan}}}$ for the full results; while the right plot gives the median of the 100 sets of parameter for each input set parameters in the top sector. The axes are the estimate and the full calculation and the black line is $y=x$.
\label{EDMscan}}
\end{figure}

\begin{figure}[h]
\includegraphics[scale=1.8]{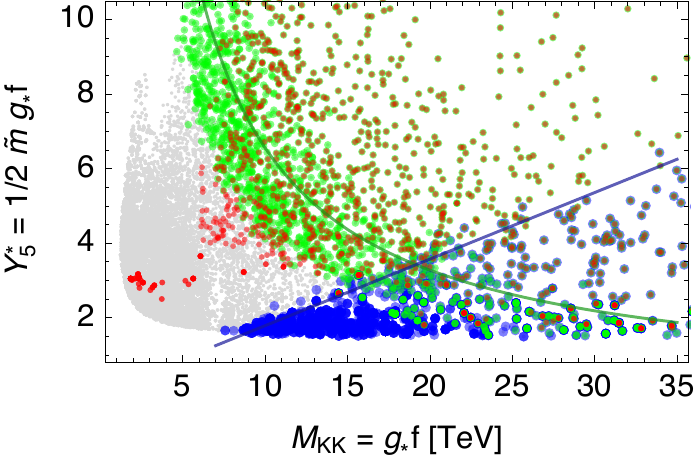}
\caption{Scatter plot of points with correct EWSB in the ($M_{KK}, Y_{5*} = \frac{1}{2} g_* \tilde{m}_d$) plane showing the impact of the flavor and EDM constraints. Gray points correspond to all points generated. Colored points indicate regions where the FCNC and EDM bounds can be satisfied. For each point we generated 100 sets of anarchic mass matrices consistent with the SM quark masses and CKM. If at least 50 iterations show (1) a small enough $C_4^K$ we color it green, (2) a small enough $C_5^K$ we color it red, and (3) a small enough neutron EDM we color it blue. The solid green (solid blue) line shows the estimates for the $C_4^K$-limit in Eq.~(\ref{KKZkaon}) (for the nEDM-limit in Eq.~(\ref{dipoleEstimate})). The allowed regions start overlapping for $M_{KK} > 18$ TeV.
\label{ScanEDMC4C5}}
\end{figure}

In Fig.~\ref{EDMscan} we compare the estimates with the full calculation of the EDM. We can see that there is a good agreement between the estimates and the full results, when we take the median of the 100 results for each set of input parameters. The analogous plots for $\text{Im}C^4_K$ and $\text{Im}C^5_K$ show a similar agreement.  Thus, our estimates give the correct typical bound on anarchic flavor in the CTH model. As usual for anarchic flavor, there is a big spread in the bounds on specific models which depends on the arbitrary choice of the distribution of the IR masses. Thus specific models could still be viable in the regions excluded in our estimates and vice versa.

\begin{figure}[h]
\includegraphics[scale=0.6]{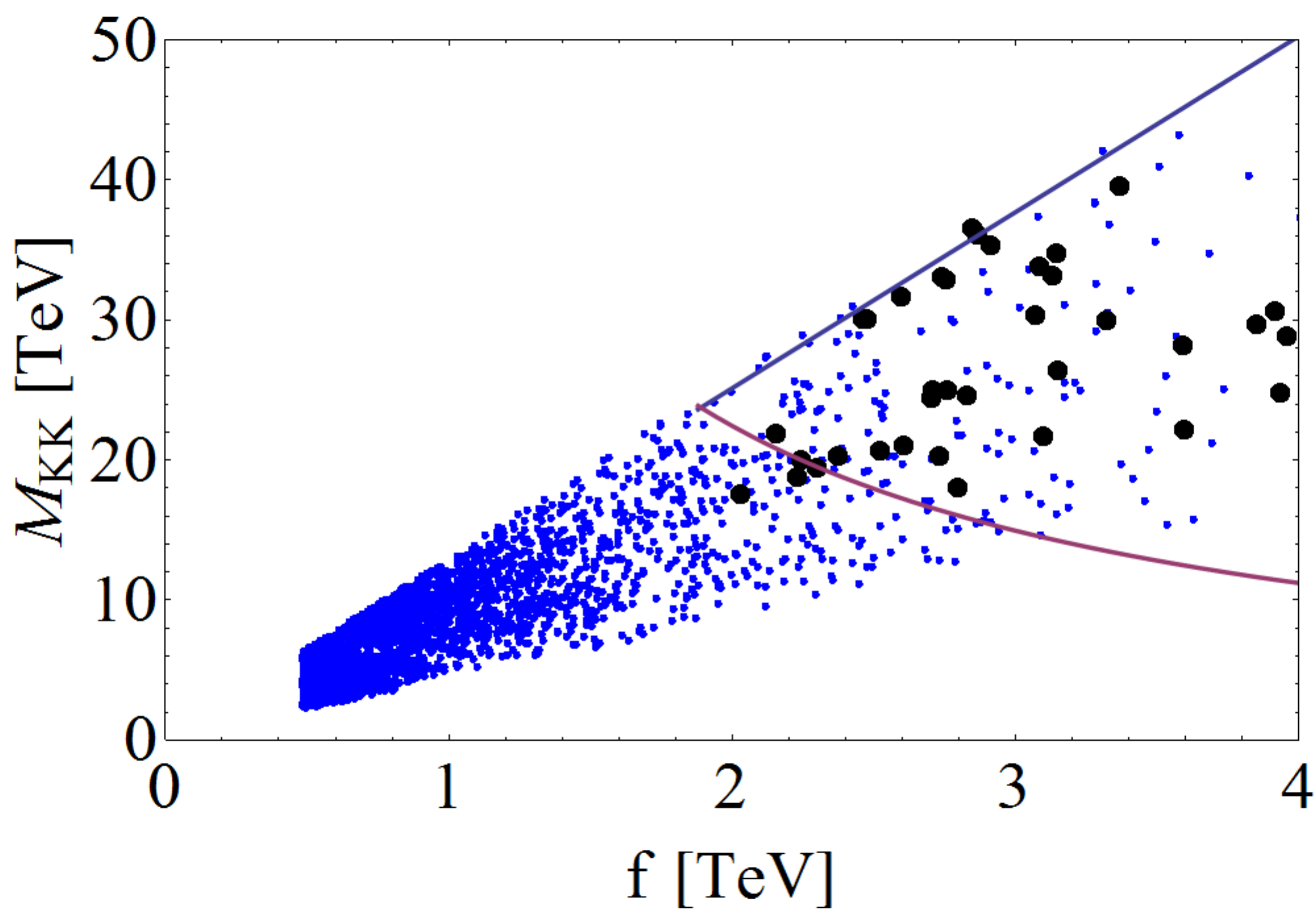}
\caption{Scatter plot of points producing proper EWSB in the ($f$,$M_{KK}$) plane. The blue line corresponds to $m_{KK}=4\pi f$ and the red line to $M_{KK}=\frac{(6.7\text{ TeV})^2}{f}$ corresponding to the combined estimate from flavor bounds. For each blue point we generated 100 sets of flavor parameters consistent with the SM values of the quark masses, CKM angles and Jarlskog invariant. The black points correspond to those for which at least half of the generated sets of flavor parameters pass all the flavor bounds.
\label{ScanSummary}}
\end{figure}

The results of the flavor scan are summarized in Figs.~\ref{ScanEDMC4C5} and \ref{ScanSummary} as a scatter plots in the ($M_{KK},Y_{5*}$)  and ($f$,$M_{KK}$) planes. The correct Higgs potential is generated for all the points shown, and the quark masses, CKM angles, as well as the Jarlskog invariant are all close to their experimental values. Fig.~\ref{ScanEDMC4C5} shows the effect of the flavor and EDM constraints, while Fig.~\ref{ScanSummary} illustrates that once the various flavor constraints are imposed, only points for which
\begin{equation}
\frac{(6.7\text{ TeV})^2}{f}<M_{KK}<4\pi f
\end{equation}
(or very close to the boundary of this region) satisfy all bounds, as expected from our estimates.

\section{Conclusions \label{sec:Conclusions}}

We have investigated the leading flavor bounds on composite twin Higgs models, and found that the tension between flavor constraints and the tuning in the Higgs potential (generically present in composite Higgs models with anarchic flavor) can be significantly reduced by considering a composite twin Higgs.  Flavor bounds imply at least few per-mille tuning for composite Higgs models, which are relaxed to the percent level for composite twin Higgs models. The reason behind this improvement is that the composite top and spin-1 partners can be raised to higher masses  in twin Higgs models (at the price of increasing their interaction strength $g_*$) without significantly affecting the Higgs potential. Instead the largest contributions are canceled by the twin partners. We have examined in detail the two leading sources of bounds: $\Delta F=2$ four-Fermi operators from KK gluon and KK-Z exchange, and dipole operators contributing to the electric dipole moment of the neutron from loops of KK fermions. We presented a simple estimate for these two quantities and the resulting bounds on the parameters of the model and verified the validity of these estimates by performing a flavor scan. This was done by generating 100 flavor models each reproducing the basic SM flavor structure for 30000 different input points with correct EWSB, and then comparing the average flavor bounds to those obtained from the estimates, resulting good agreement between the two. The best scenario corresponds to global symmetry breaking scale $f\sim 2-3$ TeV, with the interaction strength among the composites pushed close to the strong coupling limit $g_* \sim 4\pi$ (and hence generic KK modes is the 10-30 TeV range). This case satisfies all flavor bounds in the quark sector and implies a percent-level tuning in the Higgs potential. Our flavor bounds and their connections to the tuning apply both to identical and fraternal CTH models.

\section*{Acknowledgments}

We thank Kaustubh Agashe, Martin Beneke, Roberto Contino, Roni Harnik, Sebastian J\"ager, Paul Moch, Riccardo Rattazzi, Yael Shadmi, Philip Tanedo, Sal Lombardo, Emmanuel Stamou, David Straub, and Andrea Wulzer for useful discussions. We thank the CERN Theory Group for hosting the CERN-CKC Theory Institute on Neutral Naturalness where this work was initiated. C.C., M.G. and A.W. thank MIAPP in Munich for its hospitality while this work was in progress. C.C. thanks the Aspen Center for Physics for its hospitality while this work was in progress. C.C. is supported in part by the NSF grant PHY-1316222.
\newpage
\section*{Appendix}
\appendix
\section{The Quarks of the CTH: the 8-1-28 model \label{Appendix:reps}}

The SM quarks in the CTH are embedded in $\Psi_8, \Psi_1, \Psi_{28}$ bulk multiplets. The components of these multiplets, decomposed into $SU(2)_L\times U(1)_R \times SU(2)_L^m \times U(1)_R^m$ are:
\begin{eqnarray}
 \Psi_{\mathbf{8}}&=&\left\{ \begin{array}{cl} \tilde{q}_L(-,+)& ({\bf 2},1/2,{\bf 1},0) \\ q_L(+,+)&({\bf 2},-1/2,{\bf 1},0)\\ q^{m^\prime}_L(-,+)& ({\bf 1},0,{\bf 2},-1/2)\\ \tilde{q}^{m^\prime}_L(-,+)& ({\bf 1},0,{\bf 2},1/2) \end{array} \right\}_{\left({\bf 3},\frac{2}{3},{\bf 1},0\right)}~
  \Psi_{\mathbf{1}}=\left\{\begin{array}{c} u_R(+,+) ~ ({\bf 1},0,{\bf 1},0) \end{array}\right\}_{\left({\bf 3},\frac{2}{3},{\bf 1},0\right)}\nonumber \\
\Psi_{\mathbf{28}}&=&\left\{\begin{array}{c}  \begin{array}{cc} \left(\begin{array}{c} ~\\q^3_R(-,+)~({\bf 3},0,{\bf 1},0)\\~ \end{array}\right) & \left(\begin{array}{c} U_R^{5/3}(-,+) ~({\bf 1},1,{\bf 1},0)\\ U^\prime_R(-,+)~({\bf 1},0,{\bf 1},0) \\d_R(+,+)~({\bf 1},-1,{\bf 1},0)   \end{array}\right) \\ \left(\begin{array}{c} ~\\q^{3m'}_R(-,+)~({\bf 1},0,{\bf 3},0)\\~\end{array}\right) & \left(\begin{array}{c} d^{m\prime}_R(-,+)~({\bf 1},0,{\bf 1},-1)\\ U^{m\prime}_R(-,+)~({\bf 1},0,{\bf 1},0) \\U_R^{m5/3}(-,+) ~({\bf 1},0,{\bf 1},1)\end{array} \right)\end{array} \\ q^{16}_R(-,+)~({\bf 2},\pm 1/2,{\bf 2},\pm 1/2) \end{array}\right\}_{\left({\bf 3},\frac{2}{3},{\bf 1},0\right)}\ , \label{fermiondecom}
\end{eqnarray}
where for each component the $SU(2)_L\times U(1)_R \times SU(2)_L^m \times U(1)_R^m$ representations are given in the adjacent parentheses, and for the entire multiplet, the subscript gives the $SU(3)\times U(1)_X \times SU(3)^m \times U(1)_X^m$ representations. For each component we give the UV and IR brane b.c. for one of the chiralities while the opposite chirality has opposite b.c.

The mirror quarks are embedded in the $\Psi^m_8, \Psi^m_1, \Psi^m_{28}$ bulk multiplets with $\left({\bf 1},0,{\bf 3},\frac{2}{3}\right)$ quantum numbers under $SU(3)\times U(1)_X \times SU(3)^m \times U(1)_X^m$. These multiplets decompose similarly to the SM multiplets under $SU(2)_L\times U(1)_R \times SU(2)_L^m \times U(1)_R^m$. In the mirror sector, the IR b.c. are similar to the SM due to the IR $SO(7) \times Z_2$. In the UV, however,
the $Z_2$ exchanges SM and mirror gauge symmetries, and so the b.c. for the mirror multiplets can be read from Eq.~(\ref{fermiondecom}) with:
\begin{equation}
SU(2)_L\times U(1)_R \leftrightarrow SU(2)^m_L\times U(1)^m_R
\end{equation}
i.e. the b.c. for the mirror multiplets are:
\begin{eqnarray}
 \Psi_{\mathbf{8}}&=&\left\{ \begin{array}{cl} \tilde{q}^{\prime}_L(-,+)& ({\bf 2},1/2,{\bf 1},0) \\ q^{\prime}_L(-,+)&({\bf 2},-1/2,{\bf 1},0)\\ q^{m}_L(+,+)& ({\bf 1},0,{\bf 2},-1/2)\\ \tilde{q}^{m}_L(-,+)& ({\bf 1},0,{\bf 2},1/2) \end{array} \right\}_{\left({\bf 1},0,{\bf 3},\frac{2}{3}\right)}~
  \Psi_{\mathbf{1}}=\left\{\begin{array}{c} u^m_R(+,+) ~ ({\bf 1},0,{\bf 1},0) \end{array}\right\}_{\left({\bf 1},0,{\bf 3},\frac{2}{3}\right)} \nonumber \\
 \Psi_{\mathbf{28}}&=&\left\{\begin{array}{c}  \begin{array}{cc} \left(\begin{array}{c} ~\\q^{3\prime}_R(-,+)~({\bf 3},0,{\bf 1},0)\\~ \end{array}\right) & \left(\begin{array}{c} U_R^{5/3 \prime}(-,+) ~({\bf 1},1,{\bf 1},0)\\ U^{\prime \prime}_R(-,+)~({\bf 1},0,{\bf 1},0) \\d^{\prime}_R(-,+)~({\bf 1},-1,{\bf 1},0)   \end{array}\right) \\ \left(\begin{array}{c} ~\\q^{3m}_R(-,+)~({\bf 1},0,{\bf 3},0)\\~\end{array}\right) & \left(\begin{array}{c} d^{m}_R(+,+)~({\bf 1},0,{\bf 1},-1)\\ U^{m \prime \prime}_R(-,+)~({\bf 1},0,{\bf 1},0) \\U_R^{m5/3 \prime}(-,+) ~({\bf 1},0,{\bf 1},1)\end{array} \right)\end{array} \\ q^{m 16}_R(-,+)~({\bf 2},\pm 1/2,{\bf 2},\pm 1/2) \end{array}\right\}_{\left({\bf 1},0,{\bf 3},\frac{2}{3}\right)}\ , 
\end{eqnarray}
\section{Fraternal Boundary Conditions \label{Appendix:FBC}}

In the model above the low energy 4D theory contains light mirror fermions. However, it is trivial to eliminate them and the mirror photon from the spectrum by assigning  Dirichlet b.c. to them on the UV brane - and explicitly breaking $Z_2$ in the light sector.
In this case the b.c. for the light mirror quark multiplets read:
\begin{eqnarray}
 \Psi_{\mathbf{8}}&=&\left\{ \begin{array}{cl} \tilde{q}^{\prime}_L(-,+)& ({\bf 2},1/2,{\bf 1},0) \\ q^{\prime}_L(-,+)&({\bf 2},-1/2,{\bf 1},0)\\ q^{m}_L(-,+)& ({\bf 1},0,{\bf 2},-1/2)\\ \tilde{q}^{m}_L(-,+)& ({\bf 1},0,{\bf 2},1/2) \end{array} \right\}_{\left({\bf 1},0,{\bf 3},\frac{2}{3}\right)}~
  \Psi_{\mathbf{1}}=\left\{\begin{array}{c} u^m_R(-,+) ~ ({\bf 1},0,{\bf 1},0) \end{array}\right\}_{\left({\bf 1},0,{\bf 3},\frac{2}{3}\right)}\nonumber \\
\Psi_{\mathbf{28}}&=&\left\{\begin{array}{c}  \begin{array}{cc} \left(\begin{array}{c} ~\\q^{3\prime}_R(-,+)~({\bf 3},0,{\bf 1},0)\\~ \end{array}\right) & \left(\begin{array}{c} U_R^{5/3 \prime}(-,+) ~({\bf 1},1,{\bf 1},0)\\ U^{\prime \prime}_R(-,+)~({\bf 1},0,{\bf 1},0) \\d^{\prime}_R(-,+)~({\bf 1},-1,{\bf 1},0)   \end{array}\right) \\ \left(\begin{array}{c} ~\\q^{3m}_R(-,+)~({\bf 1},0,{\bf 3},0)\\~\end{array}\right) & \left(\begin{array}{c} d^{m}_R(-,+)~({\bf 1},0,{\bf 1},-1)\\ U^{m \prime \prime}_R(-,+)~({\bf 1},0,{\bf 1},0) \\U_R^{m5/3 \prime}(-,+) ~({\bf 1},0,{\bf 1},1)\end{array} \right)\end{array} \\ q^{m 16}_R(-,+)~({\bf 2},\pm 1/2,{\bf 2},\pm 1/2) \end{array}\right\}_{\left({\bf 1},0,{\bf 3},\frac{2}{3}\right)}\ , 
\end{eqnarray}
The mirror leptons (except for the mirror tau and mirror tau neutrino) can be eliminated from the spectrum in a similar manner. In this way our composite framework can also UV complete the fraternal twin Higgs of~\cite{Craig:2015pha}.

\section{The Higgs Potential in the CTH \label{Appendix:Higgs}}

The spectral functions in the top and mirror top sectors are:

\begin{eqnarray}
\rho_t(p^2) &=& 1 + f_t(p^2) \ \sin^2\left(\frac{h}{f}\right) \\
\rho_{t^m}(p^2) &=& 1 + f_t(p^2) \ \cos^2\left(\frac{h}{f}\right)
\end{eqnarray}
where
\begin{equation}
f_t = -\frac{\frac{1}{2} C_{-1} \left(\frac{R'}{R}\right)^{2c_u-2c_q}\wt{m}^2_u}{\left(C_{-8}S_{1} + C_{-1}S_{8}\left(\frac{R'}{R}\right)^{2c_u-2c_q} \wt{m}^2_u\right) S_{-8}}.
\end{equation}
\ \quad \\
The fermion eigenfunctions $C_{c}$ and $S_{c}$ are defined as
\begin{eqnarray}
(kz)^{c+2}C_{c}\left(z,p\right)&=&\frac{\pi p}{2k}(kz)^\frac{5}{2}\left[J_{c+\frac{1}{2}}\left(\frac{p}{k}\right)Y_{c-\frac{1}{2}}\left(z p\right)-Y_{c+\frac{1}{2}}\left(\frac{p}{k}\right)J_{c-\frac{1}{2}}\left(z p\right)\right] \\
(kz)^{c+2}S_{c}\left(z,p\right)&=&\frac{\pi p}{2k}(kz)^\frac{5}{2}\left[J_{\frac{1}{2}-c}\left(\frac{p}{k}\right)Y_{\frac{1}{2}-c}\left(z p\right)-Y_{\frac{1}{2}-c}\left(\frac{p}{k}\right)J_{\frac{1}{2}-c}\left(z p\right)\right],
\end{eqnarray}
and are the analogues of trigonometric functions for AdS space, in the sense that $C_{c}\left(R,p\right)=1$, ${\partial}_z C_{c}\left(R,p\right)=0$, $S_{c}\left(R,p\right)=0$, and ${\partial}_z S_{c}\left(R,p\right)=p$.

The top and top mirror masses are
\begin{equation}
m_{t}\simeq m_{t0}\sin\frac{h}{f}~,~m_{tm}\simeq m_{t0}\cos\frac{h}{f},
\end{equation}
where
\begin{equation}
m_{t0}=\sqrt{p^2f_t}|_{p \to 0}
\end{equation}
The Higgs potential is then \cite{Falkowski:2006vi}:
\begin{equation}
V_{eff}(h)=\frac{-4N_c}{(4\pi)^2}\int_0^\infty dp \, p^3 \log(\rho_t[-p^2]\rho_{tm}[-p^2]).
\label{apHiggspotential}\end{equation}
The integral in Eq.~(\ref{apHiggspotential}) can be expanded as
\begin{equation}
V_{eff}(h)=-\alpha_2 \sin^2\frac{h}{f} - \alpha_4 \sin^4 \frac{h}{f} -n_t \sin^4 \frac{h}{f}\log\frac{2m^2_{t0} \sin^2\frac{h}{f}}{\Lambda^2} + (\sin \to \cos),
\label{App:CTHpotentialFull}
\end{equation}
where $\alpha_2,\alpha_4,n_t$ are given by
\begin{eqnarray}
\alpha_2&=& \frac{3}{8\pi^2}\int \limits_{0}^\infty dp^2~p^2 f_t \\
\alpha_4&=& \frac{3}{16\pi^2}\int \limits_{0}^\infty dp^2~p^2 \left(-f^2_t + \frac{m_{t0}^4}{\Lambda^4\sinh^2(p^2/\Lambda^2)}\right) - \frac{3}{2}m_{t0}^4  \\
n_t&=& \frac{3}{16\pi^2}m_{t0}^4,
\end{eqnarray}
where $\Lambda$ is an arbitrary IR regulator.
In this paper we neglect the term
\begin{equation}
n_t\left( \sin^4\frac{h}{f}\log \left(\sin^2\frac{h}{f}\right)+\cos^4\frac{h}{f}\log\left( \cos^2\frac{h}{f}\right)\right)
\end{equation}
Which results in a shift of roughly 10\% in the Higgs mass and VEV. We note that the theoretical error in the Higgs sector is typically larger than that for large values of $g_*$. The Higgs potential can then be written as:
\begin{equation}
V_{eff}(h)=2\left(-\alpha _4-n_t \log\frac{2m^2_{t0}}{\Lambda^2}\right)\sin^2\frac{h}{f}\cos^2\frac{h}{f} \equiv -\alpha \sin^2\frac{h}{f}\cos^2\frac{h}{f}
\end{equation}

In the CTH, we don't use $\sin\frac{h}{f}$ as our small parameter, because every term with a $\sin^n \frac{h}{f}$ is accompanied by a $\cos^n \frac{h}{f}$. Instead, we define:
\begin{equation}
f_t  \equiv f^\prime_t \frac{y^2_t}{2 g_*^2}
\end{equation}
and thus we can use $x=\frac{y^2_t}{2 g_*^2} \sin^2\frac{h}{f}$ as our small parameter in SM sector and $x^m=\frac{y^2_t}{2 g_*^2} \cos^2\frac{h}{f}$ in the mirror sector and get the same expansion. All the higher terms are then suppressed by powers of this small parameter.
The low energy behavior of $f^\prime_t$ is:
\begin{equation}
f^\prime_t = \frac{M^2_{KK}}{p^2} \label{eq:fprime}
\end{equation}
and for $p\gg M_{KK}$, $f^\prime_t\sim e^{-\frac{4p}{M_{KK}}}$. Thus, there is only one dimension-full parameter in $f^\prime_t$ with ${\cal O}(1)$ factors. We can estimate that the $n^{th}$ term in the expansion of
\begin{equation}
V_{eff}\propto \int dt~t\log (1+f^\prime_t x)
\end{equation}
is proportional to
\begin{equation}
M_{KK}^4 x^n
\end{equation}
We can see that this estimate works for the above calculation of the $\alpha_2$ and $\alpha_4$, and we assume that it holds for higher orders too so that they can be safely dropped.

\section{$Z_2$ - breaking via hypercharge \label{Appendix:Z2b}}

In section \ref{sec:CTH} we have shown that the $Z_2$ conserving contribution of the top and the mirror top to the Higgs potential can give the right EWSB, when accompanied by an additional $Z_2$ breaking contribution.

There are various possible ways to generate this term \cite{GT,Barbieri:2015lqa,Low:2015nqa}. In this paper we assume that the $Z_2$ breaking originates from a mismatch between the gauge couplings $g_X$ and $g^m_X$ of ${U(1)}_X$ and ${U(1)}^m_X$. This in turn results in a shift of the $Z^m$ mass relative to the $Z_2$ symmetric value $\left(f/v\right)m_Z$.

The gauge coupling of the SM hypercharge is defined by:
\begin{equation}
 \frac{1}{g^{\prime 2}}= \log \frac{R'}{R} \left(\frac{1}{g_*^2}+\frac{1}{g_{X*}^2}\right) \approx \frac{1}{g_{X*}^2} \log \frac{R'}{R}
\end{equation}
and similarly for the mirror hypercharge.

There are two distinct contributions to the Higgs potential from the mismatch in the hypercharge gauge couplings that generate a $\beta \sin^2 \frac{h}{f}$ term:
\begin{enumerate}
\item The hypercharge gauge loops contributing directly to the Higgs potential.
\item A detuning between the top and the mirror top bulk masses due to hypercharge gauge loops. This in turn generates a $Z_2$ breaking term in the Higgs potential at two-loop level.
\end{enumerate}

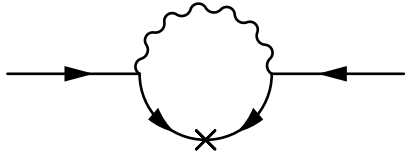
\begin{figure}[h]
\begin{center}
	    \begin{fmffile}{bulkmass}
	        \begin{fmfgraph*}(150,100)
                 \fmfstraight
	        \fmfleft{i1}
	        \fmfright{o1}
                 \fmf{fermion,tension=2}{i1,v1}
                 \fmf{fermion,tension=2}{o1,v2}
                 \fmf{wiggly,left,tag=1}{v1,v2}
                 \fmf{phantom,left,tag=2}{v2,v1}
                 \fmfposition
                 \fmfipath{p[]}
                 \fmfiset{p1}{vpath1(__v1,__v2)}
                 \fmfiset{p2}{vpath2(__v2,__v1)}
                 \fmfi{fermion}{subpath (0,length(p2)/2) of p2}
                 \fmfi{fermion}{subpath (length(p2),length(p2)/2) of p2}
               \fmfiv{d.sh=cross,d.ang=90,d.siz=5thick}{point length(p2)/2 of p2}
	        \end{fmfgraph*}
	    \end{fmffile}
\end{center}
\vspace*{-1cm}
\caption{A hypercharge loop contribution to the fermion bulk mass. \label{bulkmass}}
\end{figure}

The first contribution can be estimated to be:
\begin{equation}
\beta_1 \approx \frac{3}{128 \pi^2} (g^{\prime2}-g^{\prime2}_m) g_*^2f^4 \approx  \frac{3}{128 \pi^2} \frac{(g^2_{X*}-g^{m2}_{X*})}{\log \frac{R'}{R}} g_*^2f^4 \sim {\delta}_{g^2_{X*}} \alpha_0, \label{firstbeta}
\end{equation}
With ${\delta}_{g^2_{X*}}\equiv (g^2_{X*}-g^{m2}_{X*})/g^2_{X*}$.
For $g_{X*}>g^m_{*X}$ this contribution has the right sign to generate $v<f$. The second contribution is due to hypercharge gauge loops in the bulk shown in Fig.~\ref{bulkmass}.
Evaluating these loops, we can estimate the shift in the bulk mass:
\begin{equation}
 \Delta c \sim c \frac{q_X^2 \Delta {g_X}^2}{24\pi^3} \int \frac{d^5p}{p^4}
\end{equation}
This integral is linearly divergent and therefore strongly depends on the dynamics at the 5D cutoff. We can estimate $\Delta c$ by using a cutoff $\Lambda$, estimated in turn from the strongest coupling in the theory - the SO(8) bulk gauge coupling. The resulting $Z_2$ mismatch in the bulk mass for $g_*\sim 4\pi$ is
\begin{equation}
\Delta c \sim c q_X^2 \frac{\Delta{g_*'}^2}{g_*^2} \sim 0.005 {\delta}_{g^2_{X*}},
\end{equation}
where $q_X$ is the charge of the fermions under $U(1)_X$, and $g_*'=\frac{g_X}{\sqrt{R}}$. We can now calculate the effect of this mismatch on the Higgs potential. An estimate for this term is:
\begin{equation}
\beta_2 =\alpha_2(c)-\alpha^m_2(c+\Delta c)\sim \Delta c \frac{d  \alpha_{2}(c)}{d c},
\end{equation}
where $\alpha_2(c)$ is the term in the Higgs potential that is quadratically sensitive to $M_{KK}$ (which is exactly canceled by $\alpha_2^m$ in the $Z_2$ symmetric limit). The NDA for this term for $g_*\sim 4\pi$ is:
\begin{equation}
\alpha_2\sim \frac{3}{16 \pi^2}y_t^2 g_*^2 f^4 \sim 50 \alpha_0\ ,
\end{equation}
while the derivative can be estimated as
\begin{equation}
\frac{d \alpha_2}{dc}\sim 2 \alpha_2 \frac{d \log y_t}{dc}
\end{equation}
For the points that give the right $\alpha$, we find that $\frac{d \log y_t}{dc}\sim 3 $, so that
\begin{equation}
\beta_2 \sim 2\Delta c \, \alpha_{2}  \frac{d \log y_t}{dc} \sim  {\delta}_{g^2_{X*}} \alpha_0
\label{secondbeta}\end{equation}
Hence, both contributions to the Higgs potential in Eqs.(~\ref{firstbeta}, \ref{secondbeta}) have the right size to to generate the correct EWSB, according to Eq.~(\ref{alphabetaf}). The overall $\beta$ is:
\begin{equation}
\beta \sim {\delta}_{g^2_{X*}} \alpha_0
\end{equation}
so that each value of $f$ requires a different value of $g^{m}_{X*}$. For the calculation of the tuning in Eq.~(\ref{tuning}), we assume  ${\delta}_{g^2_{X*}}\sim {\cal O}(1)$, so that $\Delta = \Delta_{\alpha}$.

\section{The Tuning in the CH \label{Appendix:TuneCH}}

In this section we elaborate on the tuning in the CH, as presented in Eq.~(\ref{eq:Tuning_CH}). The NDA for the tuning depends on the particular SO(5) representations in which the top quark is embedded~\cite{Panico:2012uw}.
Specifically, we have calculated the tuning in the doubly-tuned CH model with two adjoints and a fundamental of SO$(5)$, studied previously in
\cite{CFW} and \cite{Carena:2007ua}. In this model, the top and gauge contributions generate a Higgs potential of the form:
\begin{equation}
V(h)=-\alpha \sin^2\frac{h}{f} + \beta \sin^4\frac{h}{f},
\end{equation}
with $\alpha$ and $\beta$ depending on the bulk masses $(c_q, c_u, c_d)$, the IR masses $(m_u, M_u, m_d)$, and the UV/IR hierarchy $R'/R$. Both $\alpha$ and $\beta$ scale as $f^4$ times a function of $\left(R'/R\right)$. Consequently:
\begin{equation}
v \equiv f \sin \left(\left<h\right>/f \right) \sim f \times F(R'/R, c_q, c_u, c_d, \tilde{m}_u, \tilde{M}_u, \tilde{m}_d),
\end{equation}
with $f=2/g_* R'$.
The top mass is given by:
\begin{equation}
m_t = \frac{g_* v}{2\sqrt{2}} \frac{f_q f_{-u} \left| \hat{m}_u - \hat{M}_u\right|}
{\sqrt{\left(1+f^2_q f^{-2}_u m^2_u+f^2_q f^{-2}_d \hat{m}^2_d\right)\left(1+f^2_{-u} f^{-2}_{-q} \hat{M}^2_u\right)}},
\end{equation}
where $\hat{m}_{u,d}={\left(R'/R\right)}^{c_{u,d}-c_q}\tilde{m}_{u,d}$, $\hat{M}_{u}={\left(R'/R\right)}^{c_{u}-c_q}\tilde{M}_{u}$, and $f_c$ is the RS flavor function. The W mass is
\begin{equation}
m_W = \frac{gv}{2}, \quad \frac{1}{g^2} \equiv \frac{1}{{g_*}^2} \left(1+r^2 \right) \log{\left(\frac{R'}{R}\right)},
\end{equation}
with $r$ the UV kinetic term for $SU(2)_L$.

\begin{figure}[h]
\includegraphics[scale=0.5]{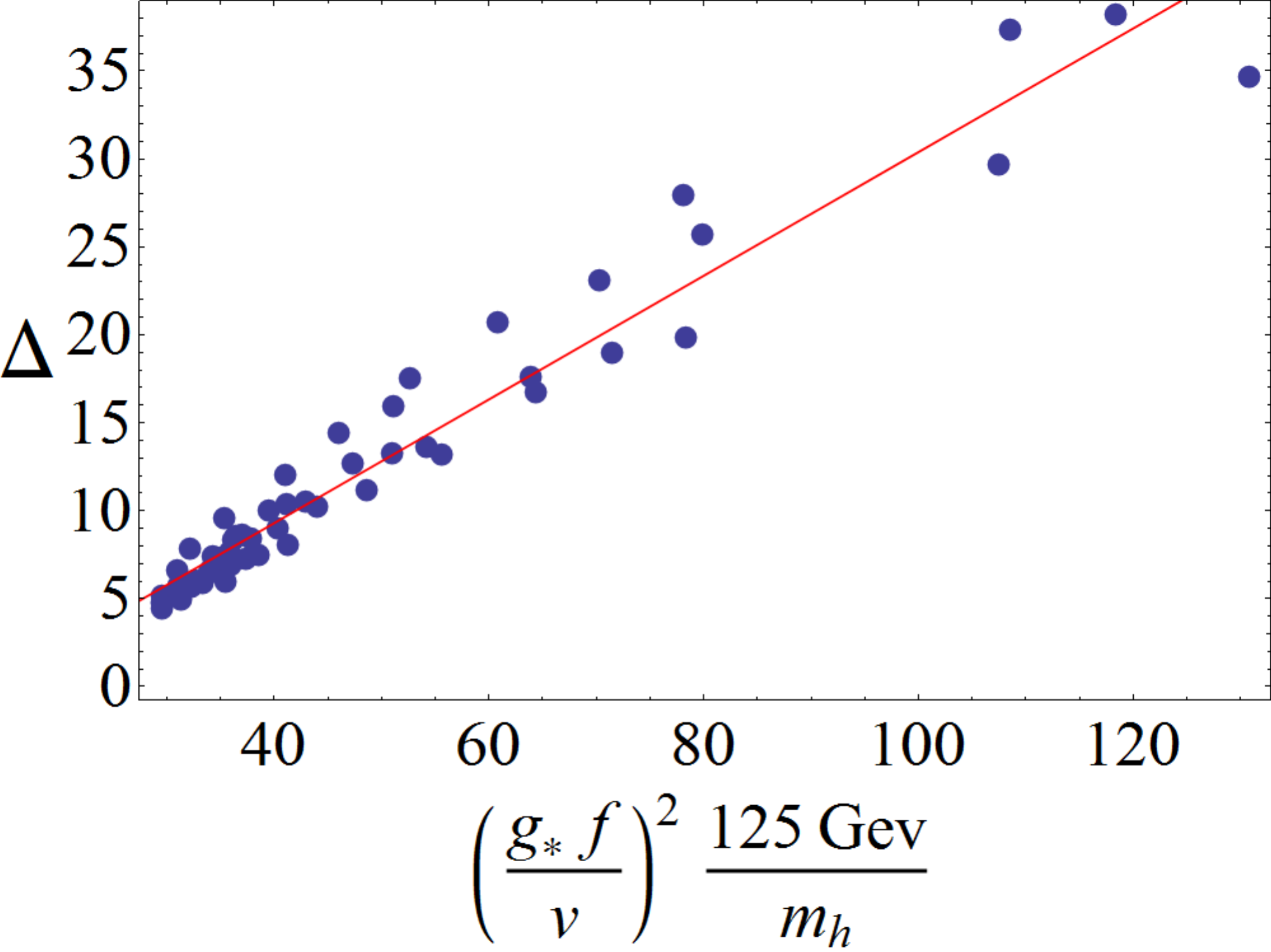}
\caption{Scatter plot for the tuning $\Delta$ in the CH model vs. ${\left(g_* f / v\right)}^2 \times \left(125GeV / m_h \right)$. The range for the scan is
$R'/R \in [1,1.4] \times 10^{16}, \  c_u \in [-0.5,-0.2], \ c_d=-0.55,\
 c_q \in [0.36, 0.4], \ m_u \in [0.5, 4], \  M_u=0, m_d=2.$
\label{fig:Tuning_CH}}
\end{figure}

We scan for points in the parameter space that reproduce $v, \ m_W$ and $m_t$ in the following way. For each point $(R'/R, c_q, c_u, c_d, \tilde{m}_u, \tilde{M}_u, \tilde{m}_d)$, we set $g_*$ so that $m_t = m^{exp}_t$. We then set the IR gauge kinetic term $r$ such that $m_W$ is reproduced. With $g_*$ and $R'/R$ held fixed, the VEV $v \equiv f \sin \left(\left<h\right>/f\right) $ scales like $R^{-1}$, so we can set $R$ to reproduce the correct electroweak symmetry breaking.  The resulting Higgs mass is in the 100-200 GeV range, with additional tuning required to obtain its precise value of 125 GeV. Ignoring this extra tuning for now, we define the tuning in this model analogously to (\ref{tuning}):
\begin{equation}
\Delta = \max \left|\frac{d\log v}{d\log p_i}\right|,
\label{tuning_CH}
\end{equation}
with $p_i = c_q, c_u, c_d, \tilde{m}_u, \tilde{M}_u, \tilde{m}_d, g_*, r, R, f$. The results of the scan are given in Fig~\ref{fig:Tuning_CH}, were the tuning is plotted for various points reproducing $v, m_t$ and $m_W$.  We see that the tuning can be well approximated by $0.35{\left(\frac{g_* f} { v}\right)}^2 \frac{125~\text{GeV}}{ m_h }$. This dependence on $f, g_*$ and $m_h$ is similar to the ones estimated in~\cite{Panico:2012uw} for similar doubly-tuned CH models.

\section{Loop Functions in the Dipole Calculation \label{Appendix:dipole}}

The loop functions $L^\Psi_X$ (where $\Psi$ is the KK fermion and $X$ is any of $H,W,Z,W^m,Z^m$) are given in \cite{Konig:2014iqa}. In the limit $M_{\Psi} \gg M_X$, these are:
\begin{equation}
L^\Psi_{V} \approx -\frac{Q_\Psi + Q_V}{4} ~,~L^\Psi_{H} \approx -\frac{ Q_\Psi M_H^2}{M_\Psi^2}\ ,
\end{equation}
where $Q_\Psi$ and $Q_V$ are the charges of the fermion and the boson in the loop. For the calculation of the $\bar{d}d\gamma$ operator they are:
\begin{eqnarray}
X&=&H,Z,Z^m,W^m,~Q_\Psi = -\frac{1}{3},Q_V=0\\
X&=&W^+,~Q_\Psi = \frac{2}{3},Q_V=1\\
X&=&W^-,~Q_\Psi = -\frac{4}{3},Q_V=-1
\end{eqnarray}
For the $\bar{d}dG$ operator, $Q_\Psi=1,Q_V=0$ for any $X$.


\begin{thebibliography}{99}
\bibitem{Aad:2015kqa}
  G.~Aad {\it et al.} [ATLAS Collaboration],
  JHEP {\bf 1508}, 105 (2015)
  [arXiv:1505.04306 [hep-ex]].

\bibitem{CMS:2013tda}
  CMS Collaboration [CMS Collaboration],
  CMS-PAS-B2G-12-015.

\bibitem{Chacko:2005pe}
  Z.~Chacko, H.~S.~Goh and R.~Harnik,
  Phys.\ Rev.\ Lett.\  {\bf 96} (2006) 231802
  [hep-ph/0506256].

\bibitem{Chacko:2005un}
  Z.~Chacko, H.~S.~Goh and R.~Harnik,
  JHEP {\bf 0601} (2006) 108
  [hep-ph/0512088].
\bibitem{Craig:2013fga}
  N.~Craig and K.~Howe,
  JHEP {\bf 1403} (2014) 140
\bibitem{Craig:2014aea}
  N.~Craig, S.~Knapen and P.~Longhi,
  Phys.\ Rev.\ Lett.\  {\bf 114} (2015) 6,  061803
  [arXiv:1410.6808 [hep-ph]].

\bibitem{Craig:2014roa}
  N.~Craig, S.~Knapen and P.~Longhi,
  JHEP {\bf 1503} (2015) 106
  [arXiv:1411.7393 [hep-ph]].

\bibitem{GT}
 M.~Geller and O.~Telem,
  Phys.\ Rev.\ Lett.\  {\bf 114}, no. 19, 191801 (2015)
  [arXiv:1411.2974 [hep-ph]].

\bibitem{Barbieri:2015lqa}
  R.~Barbieri, D.~Greco, R.~Rattazzi and A.~Wulzer,
  JHEP {\bf 1508} (2015) 161
  [arXiv:1501.07803 [hep-ph]].

\bibitem{Low:2015nqa}
  M.~Low, A.~Tesi and L.~T.~Wang,
  Phys.\ Rev.\ D {\bf 91} (2015) 095012
  [arXiv:1501.07890 [hep-ph]].



\bibitem{Craig:2015pha}
  N.~Craig, A.~Katz, M.~Strassler and R.~Sundrum,
  JHEP {\bf 1507} (2015) 105
  [arXiv:1501.05310 [hep-ph]].


\bibitem{Curtin:2015fna}
  D.~Curtin and C.~B.~Verhaaren,
  arXiv:1506.06141 [hep-ph].

\bibitem{Csaki:2015fba}
  C.~Csaki, E.~Kuflik, S.~Lombardo and O.~Slone,
  Phys.\ Rev.\ D {\bf 92}, no. 7, 073008 (2015)
  [arXiv:1508.01522 [hep-ph]].


\bibitem{Curtin:2015bka}
  D.~Curtin and P.~Saraswat,
  arXiv:1509.04284 [hep-ph].

\bibitem{GeorgiKaplan}
  D.~B.~Kaplan and H.~Georgi,
  Phys.\ Lett.\ B {\bf 136}, 183 (1984);
  D.~B.~Kaplan, H.~Georgi and S.~Dimopoulos,
  Phys.\ Lett.\ B {\bf 136}, 187 (1984).

\bibitem{ACG}
  N.~Arkani-Hamed, A.~G.~Cohen and H.~Georgi,
  Phys.\ Lett.\ B {\bf 513}, 232 (2001)
  [hep-ph/0105239].


\bibitem{LH}
N.~Arkani-Hamed, A.~G.~Cohen, E.~Katz and A.~E.~Nelson,
  JHEP {\bf 0207}, 034 (2002)
  [hep-ph/0206021];
 N.~Arkani-Hamed, A.~G.~Cohen, E.~Katz, A.~E.~Nelson, T.~Gregoire and J.~G.~Wacker,
  JHEP {\bf 0208}, 021 (2002)
  [hep-ph/0206020].



\bibitem{Agashe:2004rs}
  K.~Agashe, R.~Contino and A.~Pomarol,
  Nucl.\ Phys.\ B {\bf 719} (2005) 165
  [hep-ph/0412089].

\bibitem{Agashe:2006at}
  K.~Agashe, R.~Contino, L.~Da Rold and A.~Pomarol,
  Phys.\ Lett.\ B {\bf 641} (2006) 62
  [hep-ph/0605341].

\bibitem{Contino:2006qr}
  R.~Contino, L.~Da Rold and A.~Pomarol,
  Phys.\ Rev.\ D {\bf 75} (2007) 055014
  [hep-ph/0612048].

\bibitem{Contino:2006nn}
  R.~Contino, T.~Kramer, M.~Son and R.~Sundrum,
  JHEP {\bf 0705} (2007) 074
  [hep-ph/0612180].

\bibitem{SILH}
 G.~F.~Giudice, C.~Grojean, A.~Pomarol and R.~Rattazzi,
  JHEP {\bf 0706}, 045 (2007)
  [hep-ph/0703164].



\bibitem{Panico:2015jxa}
  G.~Panico and A.~Wulzer,
  arXiv:1506.01961 [hep-ph].

\bibitem{Panico:2012uw}
  G.~Panico, M.~Redi, A.~Tesi and A.~Wulzer,
  JHEP {\bf 1303} (2013) 051
  [arXiv:1210.7114 [hep-ph]].

\bibitem{CHreviews}
For reviews see B.~Bellazzini, C.~Cs\'aki and J.~Serra,
  Eur.\ Phys.\ J.\ C {\bf 74}, no. 5, 2766 (2014)
  [arXiv:1401.2457 [hep-ph]];
 G.~Panico and A.~Wulzer,
  Lect.\ Notes Phys.\  {\bf 913}, pp. (2016)
  [arXiv:1506.01961 [hep-ph]];
 C.~Csaki, C.~Grojean and J.~Terning,
  arXiv:1512.00468 [hep-ph].


\bibitem{HuberShafi}
S.~J.~Huber and Q.~Shafi,
  Phys.\ Lett.\ B {\bf 498}, 256 (2001)
  [hep-ph/0010195].



\bibitem{Agashe:2004cp}
 K.~Agashe, G.~Perez and A.~Soni,
  Phys.\ Rev.\ Lett.\  {\bf 93}, 201804 (2004)
  [hep-ph/0406101];
 K.~Agashe, G.~Perez and A.~Soni,
  Phys.\ Rev.\ D {\bf 71} (2005) 016002
  [hep-ph/0408134].


\bibitem{CFW}
  C.~Csaki, A.~Falkowski and A.~Weiler,
  JHEP {\bf 0809}, 008 (2008)
  [arXiv:0804.1954 [hep-ph]].

\bibitem{rattazzi}
 B.~Keren-Zur, P.~Lodone, M.~Nardecchia, D.~Pappadopulo, R.~Rattazzi and L.~Vecchi,
  Nucl.\ Phys.\ B {\bf 867}, 394 (2013)
  [arXiv:1205.5803 [hep-ph]].


\bibitem{Agashe:2008uz}
  K.~Agashe, A.~Azatov and L.~Zhu,
  Phys.\ Rev.\ D {\bf 79} (2009) 056006
  [arXiv:0810.1016 [hep-ph]].


\bibitem{Casagrande:2008hr}
  S.~Casagrande, F.~Goertz, U.~Haisch, M.~Neubert and T.~Pfoh,
  JHEP {\bf 0810} (2008) 094
  [arXiv:0807.4937 [hep-ph]].

\bibitem{Blanke:2008zb}
  M.~Blanke, A.~J.~Buras, B.~Duling, S.~Gori and A.~Weiler,
  JHEP {\bf 0903} (2009) 001
  [arXiv:0809.1073 [hep-ph]].

\bibitem{Blanke:2008yr}
  M.~Blanke, A.~J.~Buras, B.~Duling, K.~Gemmler and S.~Gori,
  JHEP {\bf 0903} (2009) 108
  [arXiv:0812.3803 [hep-ph]].

\bibitem{Albrecht:2009xr}
  M.~E.~Albrecht, M.~Blanke, A.~J.~Buras, B.~Duling and K.~Gemmler,
  JHEP {\bf 0909} (2009) 064
  [arXiv:0903.2415 [hep-ph]].


\bibitem{Cacciapaglia:2007fw}
  G.~Cacciapaglia, C.~Csaki, J.~Galloway, G.~Marandella, J.~Terning and A.~Weiler,
  JHEP {\bf 0804} (2008) 006
  [arXiv:0709.1714 [hep-ph]].

\bibitem{CFW2}
 J.~Santiago,
  JHEP {\bf 0812}, 046 (2008)
  [arXiv:0806.1230 [hep-ph]];
C.~Csaki, A.~Falkowski and A.~Weiler,
  Phys.\ Rev.\ D {\bf 80}, 016001 (2009)
  [arXiv:0806.3757 [hep-ph]].


\bibitem{Csaki:2009wc}
  C.~Csaki, G.~Perez, Z.~Surujon and A.~Weiler,
  Phys.\ Rev.\ D {\bf 81} (2010) 075025
  [arXiv:0907.0474 [hep-ph]].

\bibitem{Redi:2011zi}
  M.~Redi and A.~Weiler,
  JHEP {\bf 1111} (2011) 108
  [arXiv:1106.6357 [hep-ph]].

\bibitem{Redi:2013eaa}
  M.~Redi, V.~Sanz, M.~de Vries and A.~Weiler,
  JHEP {\bf 1308} (2013) 008
  doi:10.1007/JHEP08(2013)008
  [arXiv:1305.3818, arXiv:1305.3818 [hep-ph]].


\bibitem{Fitzpatrick:2007sa}
  A.~L.~Fitzpatrick, G.~Perez and L.~Randall,
  Phys.\ Rev.\ Lett.\  {\bf 100} (2008) 171604
  [arXiv:0710.1869 [hep-ph]].

\bibitem{Barbieri:2012uh}
  R.~Barbieri, D.~Buttazzo, F.~Sala and D.~M.~Straub,
  JHEP {\bf 1207} (2012) 181
  doi:10.1007/JHEP07(2012)181
  [arXiv:1203.4218 [hep-ph]].

\bibitem{Azatov:2014lha}
  A.~Azatov, G.~Panico, G.~Perez and Y.~Soreq,
  JHEP {\bf 1412} (2014) 082
  [arXiv:1408.4525 [hep-ph]].

\bibitem{Carena:2007ua}
  M.~Carena, E.~Ponton, J.~Santiago and C.~E.~M.~Wagner,
  Phys.\ Rev.\ D {\bf 76} (2007) 035006
  [hep-ph/0701055].


\bibitem{ABP}
 K.~Agashe, A.~E.~Blechman and F.~Petriello,
  Phys.\ Rev.\ D {\bf 74}, 053011 (2006)
  [hep-ph/0606021].

\bibitem{GIP}
O.~Gedalia, G.~Isidori and G.~Perez,
  Phys.\ Lett.\ B {\bf 682}, 200 (2009)
  [arXiv:0905.3264 [hep-ph]].



\bibitem{CGTT}
 C.~Csaki, Y.~Grossman, P.~Tanedo and Y.~Tsai,
  Phys.\ Rev.\ D {\bf 83}, 073002 (2011)
  [arXiv:1004.2037 [hep-ph]].

\bibitem{Beneke1}
 M.~Beneke, P.~Dey and J.~Rohrwild,
  JHEP {\bf 1308}, 010 (2013)
  [arXiv:1209.5897 [hep-ph]].


\bibitem{Konig:2014iqa}
  M.~K\"onig, M.~Neubert and D.~M.~Straub,
  Eur.\ Phys.\ J.\ C {\bf 74}, no. 7, 2945 (2014)
  [arXiv:1403.2756 [hep-ph]].

\bibitem{MR}
P.~Moch and J.~Rohrwild,
  J.\ Phys.\ G {\bf 41}, 105005 (2014)
  [arXiv:1405.5385 [hep-ph]].






\bibitem{Beneke2}
M.~Beneke, P.~Moch and J.~Rohrwild,
  arXiv:1508.01705 [hep-ph].


\bibitem{btosgamma}
 M.~Blanke, B.~Shakya, P.~Tanedo and Y.~Tsai,
  JHEP {\bf 1208}, 038 (2012)
  [arXiv:1203.6650 [hep-ph]];
P.~Moch and J.~Rohrwild,
  Nucl.\ Phys.\ B {\bf 902}, 142 (2016)
  [arXiv:1509.04643 [hep-ph]].






\bibitem{Falkowski:2006vi}
  A.~Falkowski,
  Phys.\ Rev.\ D {\bf 75}, 025017 (2007)
  [hep-ph/0610336].
\end{thebibliography}
\end{document}